\documentclass[aps,pre,groupedaddress,secnumarabic,floatfix, longbibliography]{revtex4-1} 
\usepackage{amsmath, amssymb,amsbsy, graphicx, booktabs, array, natbib, bbold, color} 
\usepackage[bottom]{footmisc}
\usepackage[utf8]{inputenc}
\usepackage{graphicx}

\usepackage{amsmath, amssymb,amsbsy, graphicx, booktabs, array, color} 

\date{}

\newcommand{\bv}[1]{\mathbf{#1}} 
  
\newcommand{\dd}{\; \mathrm{d}}

\newcommand{\mutil}{\tilde{\mu}}

\newcommand{\resperp}{\zeta_{\perp}}   
\newcommand{\tresperp}{\zeta_{\perp,t}}
\newcommand{\tresperphat}{\hat{\zeta}_{\perp,t}}
\newcommand{\trespar}{\zeta_{\parallel,t}}
\newcommand{\frespar}{\zeta_{\parallel,fib}}
\newcommand{\fresperp}{\zeta_{\perp,fib}}
\newcommand{\fresperphat}{\hat{\zeta}_{\perp,fib}}
 
\newcommand{\respar}{\zeta_{\parallel}}

\newcommand{\vel}{\bv{v}}

\newcommand{\curlyG}{\mathcal{G}}

\newcommand{\ttail}{\mathbf{t}_{tail}} %Tangent vector to the tail
\newcommand{\tx}{{t}_{x}} %x component of the vector t_{tail}
\newcommand{\ty}{{t}_{y}} %y component of the vector t_{tail}
\newcommand{\tz}{{t}_{z}} %z component of the vector t_{tail}
\newcommand{\tfib}{\mathbf{t}_{fib}} %Tangent vector to the fibre
\newcommand{\thead}{\mathbf{t}_{h}} %Tangent vector to the head rotational symmetry axis
\newcommand{\rtail}{\mathbf{r}_{t}} %Position vector for the tail
\newcommand{\rfib}{\mathbf{r}_{fib}} %Position vector for the fibre
\newcommand{\rphead}{\mathbf{r}_{h}} %Position vector for the centre of the head
\newcommand{\rbase}{\mathbf{r}_{b}} %Position vector for the base of the tail, that is the point of connection of the fibres to the tail.
\newcommand{\Ltail}{L_{t}} %Length of tail
\newcommand{\Lfib}{L_{fib}}
\newcommand{\ahead}{a_{h}} %radius of head
\newcommand{\omegaflag}{\omega_{fl}}
\newcommand{\omegaphage}{\omega_{p}}
\newcommand{\rhot}{\rho_{t}}

\newcommand{\rhof}{\rho_{fib}}
\newcommand{\Rflag}{R_{fl}}

\newcommand{\vez}{\mathbf{e}_z}
\newcommand{\angleflag}{\alpha} %_{flag}}
\newcommand{\hgap}{h_{gap}}
\newcommand{\Omegarel}{\Omega_{rel}}

\newcommand{\bfib}{\mathbf{b}_{fib}}  
\newcommand{\dpot}{\delta}
\newcommand{\kdpot}{k\delta}

\newcommand{\Ltailhat}{\hat{L}_{t}} %Length of tail
\newcommand{\Lfibhat}{\hat{L}_{fib}}
\newcommand{\aheadhat}{\hat{a}_{h}} %radius of head

\begin{document}
		\title{Hydrodynamics of bacteriophage migration along bacterial flagella}
	\author{Panayiota Katsamba\footnote{ Current address: School of Mathematics, University of Birmingham, Edgbaston, Birmingham, B15 2TT, UK}}
	\affiliation{Department of Applied Mathematics and Theoretical Physics, University of Cambridge, Cambridge CB3 0WA, United Kingdom}
	\author{Eric Lauga}
	\email{e.lauga@damtp.cam.ac.uk}
	\affiliation{Department of Applied Mathematics and Theoretical Physics, University of Cambridge, Cambridge CB3 0WA, United Kingdom}
	
		\begin{abstract}
			Bacteriophage viruses, one of the most abundant entities in our planet, lack
			the ability to move independently. Instead, they crowd fluid environments in
			anticipation of a random encounter with a bacterium. Once they `land' on the cell body
			of their victim, they are able to eject their genetic material inside the host 
			cell.  Many phage species, however, first attach to the flagellar filaments of
			bacteria. Being immotile, these so-called flagellotropic phages still manage to
			reach the cell body for infection, and the process by which they move up the
			flagellar filament has intrigued the scientific community for  decades. In 1973, 
			Berg and Anderson (Nature, {\bf 245}, 380-382) proposed the nut-and-bolt mechanism in which, similarly to   a rotated nut that is able to  move along a bolt, the phage  wraps itself around a flagellar filament
			possessing helical grooves (due to the helical rows of flagellin molecules) and
			exploits the rotation of the flagellar filament in order to passively travel along it. 
			One of the main evidence for this mechanism is the fact that mutants of bacterial species such as  {\it{Escherichia coli}} and {\it{Salmonella~typhimurium}} that possess straight flagellar filaments  with a preserved helical groove structure can  still be infected by their relative  phages.  
			Using two distinct approaches to address the short-range  interactions between phages and flagellar filaments, we provide here a first-principle theoretical model for the nut-and-bolt
			mechanism applicable to mutants possessing straight flagellar filaments. Our model is fully analytical, is able to predict the speed of  translocation of a bacteriophage along a flagellar filament as a function of the geometry of both phage and bacterium, the rotation rate of the flagellar filament, and the handedness of the helical grooves, and is consistent with past  experimental observations.		
		\end{abstract}

		\date{\today}
	
	\maketitle

		\section{Introduction}\label{intro}	
\begin{figure}[t]%[!h]
		\includegraphics[width=0.8\columnwidth]{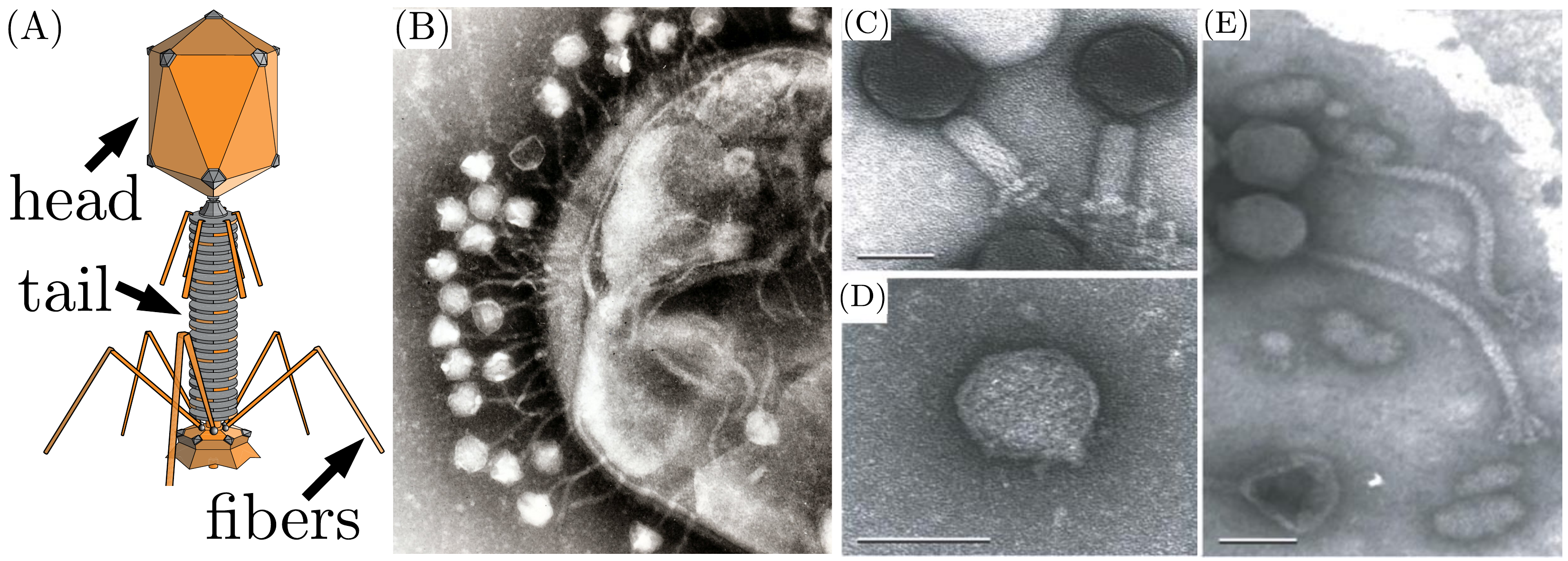}
		\caption{{\bf Bacteriophages:} (A) A typical morphology of a bacteriophage, such as the {\it Enterobacteria T4} phage (Adenosine, Wikimedia Commons); 			(B) Electron micrograph of bacteriophages attached to a bacterial cell, (Dr.~G. Beards, Wikimedia Commons); 
			(C-E) Phages can come in various shapes: (C) {\it Myoviridae}; (D) {\it Podoviridae}; (E) {\it Siphoviridae}~\cite{Suttle2005}.
			Panels C-E: Reprinted by permission from  Suttle CA, ``Viruses in the sea'', Nature, 
			{\bf 437} (356), 356-361, Copyright 2005 Springer Nature.
		}\label{Fig1}
	\end{figure}

	As big as a fraction of a micrometre, bacteriophages (in short phages), are `bacteria-eating' viruses  (illustrated in Fig.~\ref{Fig1})	 that infect bacteria and replicate within them~\cite{salmond2015}. 
	With their number estimated to be of over $10^{31}$ on the planet, phages are more abundant than every other organism on Earth  combined~\cite{Bergh1989,WommackandColwell2000,BrussowandHendrix2002,Wilhelmetal2002,Hendrix2003,HamblyandSuttle2005,Suttle2005}. 
	
	Phages have been used extensively in genetic studies~\cite{HersheyChase1952, salmond2015, BrussowandHendrix2002},
	and their future use in medicine is potentially of even greater impact. 
	The global rise in antibiotic resistance, as reported by the increasing number of multidrug-resistant bacterial infections~\cite{mechantibioticresBlairet2015}, poses one of the greatest threats to human health of our times, and phages could offer the key to resolution. 
	Indeed, phages have been killing bacteria for way longer than humanity has been 
	fighting against bacterial infections,  with as many as $10^{29}$ infections of bacterial cells by oceanic phages taking place every day~\cite{Suttle2007,Brussaardetal2008}.
	Phage therapy is an alternative to antibiotics that has been used for almost a century and offers promising solutions to tackle  antibiotic-resistant bacterial infections~\cite{Summers2001}. 
	Furthermore,	the unceasing phage-bacteria war taking place in enormous numbers offers the scientific community  great opportunities to learn. 	For example, 	the ability of phages to update their infection mechanisms in response to bacterial resistance could offer us valuable insight into updating antibiotics treatment against multi drug-resistant pathogenic bacteria~\cite{phageresistmechLabrieSamsonMoineau2010}. 
	In addition, the high selectivity of the attachment of a phage to the receptors on the bacterial cell surface and the species it infects could help  identify possible target points of particular pathogenic bacteria for drugs to attack~\cite{Rakhuba2010MechPhageAdsorption}.
	In general, extensive studies of bacteriophage infection strategies could not only reveal vulnerable points of bacteria, but may help uncover remarkable biophysical phenomena taking place at these small scales.

	Infection mechanisms can vary across the spectrum of phage species~\cite{KutterSulakvelidzeBacteriophagesBioandAppl2004,Rakhuba2010MechPhageAdsorption}. 
	Lacking the ability to move independently, phages simply crowd fluid environments and rely on a random encounter with a bacterium in order to land on its surface and accomplish infection using remarkable nanometre size machinery.
	Typically, the receptor-binding proteins located on the long tail fibres recognise and bind to the receptors of the host cell via a two-stage process called phage adsorption~\cite{Rakhuba2010MechPhageAdsorption}. 	The first stage is reversible, and is followed by irreversible attachment onto the cell surface. Subsequently, the genetic material is ejected  from their capsid-shaped head, through their tail, which is a hollow tube, into the bacterium~\cite{GraysonMolineux2007,EjectionMechanismsMolineuxPanja2013}.

\begin{figure}[!h]
		\includegraphics[width=0.8\columnwidth]{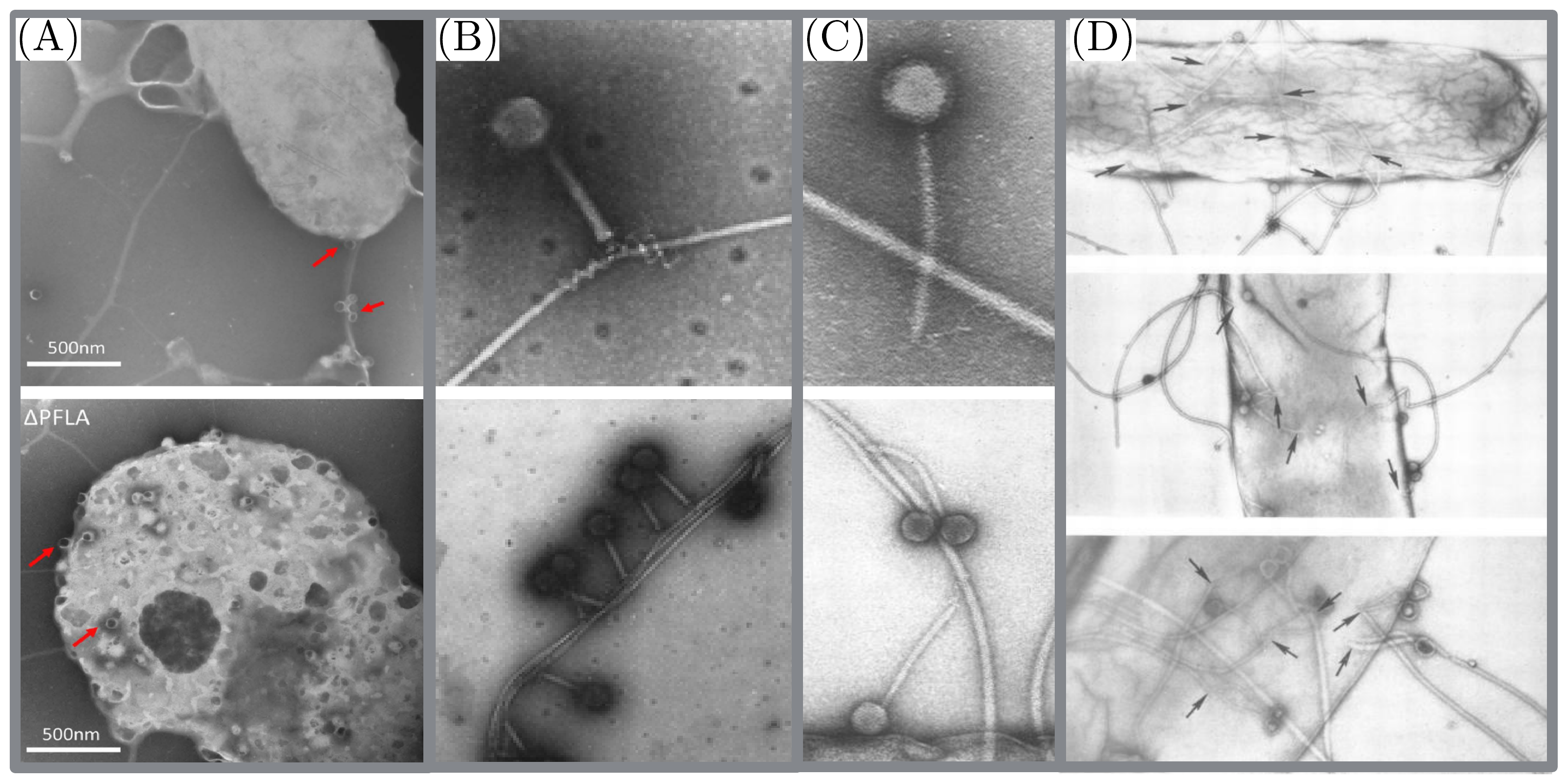}
		\caption{{\bf Flagellotropic phages:}
			(A) Attachment of phage OWB  to {\it V.~parahaemolyticus}~\cite{Zhangetal2016FlagelRotResPhage}.  Red arrows indicate phage particles. 
			(B) Phage PBS1 adsorbed 
			 to the flagellar 
			 filament of a {\it B.~Subtilis} bacterium 
with its tail fibres wrapped around the flagellar filament in a helical shape with a pitch of $35~\rm{nm}$~\cite{RaimondoLundhMartinezPBS1phageonBsubtilis1968}.   The phage 
			hexagonal head capsid measures $120~\rm{nm}$ from edge to edge~\cite{Eiserling1967}.   
			Reprinted (amended) by permission from American Society for Microbiology from  	Raimondo LM, Lundh NP, Martinez RJ, ``Primary Adsorption Site of Phage PBS1: the Flagellum of Bacillus'', J.~Virol., 1968, {\bf 2} (3), 256-264, Copyright 1968, American Society for Microbiology.
			(C)  $\chi$-phage of {\it E.~coli}~\cite{SchadeAdlerRis1967}.   The head  measures $65$ to $67.5~\rm{nm}$ between the parallel sides of the hexagon~\cite{SchadeAdlerRis1967};  
			(D) $\chi$-phage  at different times between attachment on the flagellar filament of {\it E.~coli} and reaching the base of the filament~\cite{SchadeAdlerRis1967}. 
			 Arrows point to the bases of the flagella.
			Panels C-D: Reprinted (amended) by permission from American Society for Microbiology from  Schade SZ, Adler J, Ris H, ``How Bacteriophage $\chi$ Attacks Motile Bacteria'',  J.~Virol., 1967, {\bf 1} (3), 599-609, Copyright 1967, American Society for Microbiology.
			}\label{Fig2}
	\end{figure}

	While all phages need to find themselves on the surface of the cell body for infection to take place, there is a class of phages, called flagellotropic phages, that  first attach to the flagellar filaments of bacteria. Examples include the $\chi$-phage infecting {\it Escherichia coli} ({\it E.~coli}) and {\it{Salmonella~typhimurium}} ({\it Salmonella}), the phage PBS1 infecting {\it Bacillus subtilis} ({\it B.~subtilis}) and the recently discovered phage vB\_VpaS\_OWB (for short  OWB)  infecting {\it Vibrio parahaemolyticus} ({\it V.~parahaemolyticus})~\cite{Zhangetal2016FlagelRotResPhage},  illustrated in Fig.~\ref{Fig2}. 	
	
	Given the fact that phages are themselves incapable of moving independently and that the distance they would have to traverse along the flagellar filament is   large compared to their size, they must find an active means of progressing along the flagellar filament.
	In Ref.~\cite{SchadeAdlerRis1967}, electron microscopy images of the flagellotropic $\chi$-phage, shown in Figs.~\ref{Fig2}C and D, were provided to show that the mechanism by which $\chi$-phage infects {\it E.~coli} consists of travelling along the outside of the flagellar filament until it reaches the base of the flagellar filament where it ejects its DNA.

\begin{figure}[!h]
		\includegraphics[width=0.75\columnwidth]{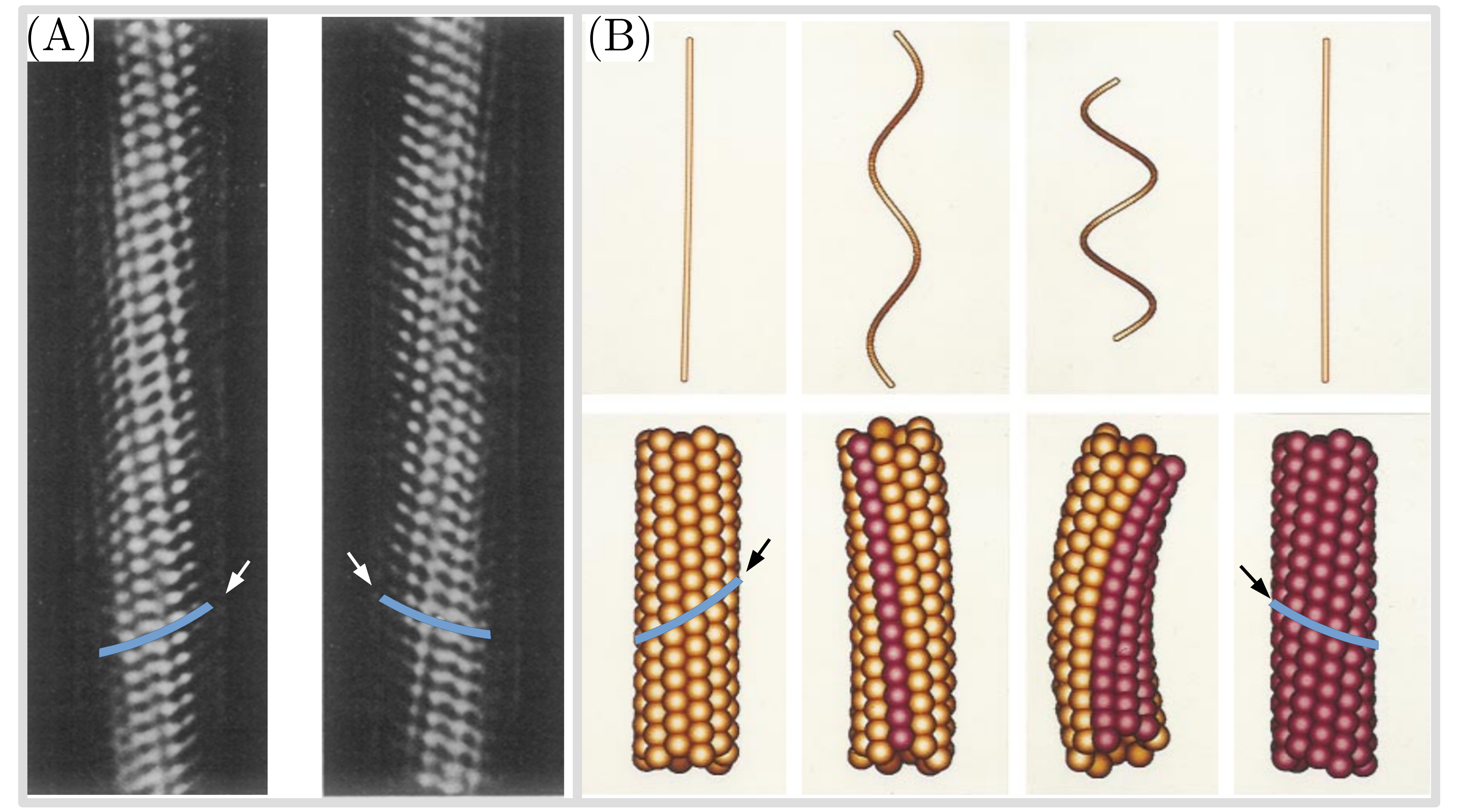}
		\caption{{\bf Bacterial flagellar filaments and polymorphism:} (A) Structure of straight flagellar filament from a mutant  of {\it Salmonella typhimurium}~\cite{BrienBennett1972}. 
			Reprinted with permission from    O'Brien EJ,  Bennett PM,  ``Structure of straight flagella from a mutant Salmonella'', J.~Mol.~Biol., {\bf 70} (1), 145-152, Copyright 1972   Elsevier.   			The L-type straight flagellar filament (left) has two types of helical grooves,  left-handed long-pitch  and right-handed short-pitch grooves, whereas  the R-type straight filament (right) has right-handed long-pitch and left-handed short-pitch grooves. 
			Examples of short-pitch grooves are marked by thin blue lines and indicated by arrows.
			(B) Schematic of some polymorphic states of the flagellar filament, from left to right: L-type straight, normal (left-handed shape), curly (right-handed) and R-type straight. The top panel shows the shape of the filaments while the bottom panel displays  the  arrangements of flagellin subunits. Examples of short-pitch grooves are marked by thin blue lines and indicated by arrows~\cite{NambaVonderviszt1997}. 	Reprinted with permission from 
			Namba K,  Vonderviszt F,
			``Molecular architecture of bacterial flagellum'', Q.~Rev.~Biophys., 1997, {\bf 30} (1), 1-65, 	Copyright 1997 Cambridge University Press. 
		}\label{Fig3}
	\end{figure}

	A possible mechanism driving the translocation of $\chi$-phage along the flagellar filament was first proposed in Berg and Anderson's    seminal paper as the `nut-and-bolt' mechanism~\cite{BergAnderson1973}. Their paper  is best known for establishing that bacteria swim by rotating their flagellar filaments. One of the supporting arguments was the proposed mechanism where the  phage plays the role of the nut and the bolt is the  flagellar filament, with the grooves between the helical rows of flagellin molecules  making up the flagellar filament  serving as the threads~\cite{BergAnderson1973} (Figs.~\ref{Fig3}A and  B). A phage would then wrap around the flagellar filament and the rotation of the latter would result in the translocation of the phage along it. 
	
	A mutant of {\it Salmonella} that has straight flagellar filaments, but possesses the same helical screw-like surface due to the arrangement of the flagellin molecules~\cite{BrienBennett1972} is non-motile due to the lack of chiral shape yet fully sensitive to $\chi$-phage~\cite{MutantSalmStraightfil1967}, i.e.~the phages  manage to get transported to the base of the flagellar filament. This is consistent with the nut-and-bolt mechanism and was used as evidence that the flagellar filament is rotating~\cite{BergAnderson1973}.

	More evidence in support of the nut-and-bolt mechanism were provided 26 years after its inception in a work studying strains of {\it Salmonella} mutants with straight flagellar filaments  whose motors alternate from rotating clockwise (CW) and counter-clockwise (CCW)~\cite{Samueletal1999Berg}. 
	The directionality of rotation is crucial to the mechanism as CCW rotation will only pull the phage toward the cell body if the phage slides along a right-handed groove. In order to test the directionality, the authors used a chemotaxis signalling protein that interacts with the flagellar motor, decreasing the CCW bias. They found that 
	strains 
	with a large CCW bias are sensitive to $\chi$-phage infection, whereas those with small CCW bias are resistant, in agreement with the proposed nut-and-bolt mechanism.

	Details of the packing of the flagellin molecules that give rise to the grooves can be found in Ref.~\cite{NambaVonderviszt1997} and examples are shown in Figs.~\ref{Fig3}A and B. It is important to note that the packing of flagellin molecules produces two overlapping sets of helical grooves, a long-pitch and a short-pitch set of grooves which are of opposite chirality~\cite{BrienBennett1972}. Once in contact with a rotating flagellar filament, it is anticipated that the phage fibres will wrap along the short-pitch grooves. Indeed, the findings of Ref.~\cite{Samueletal1999Berg} show that the directionality of phage translocation correlates with the chirality of the short-pitch grooves. %}
	
	The flagellar filaments of bacteria can take  one of the twelve distinct polymorphic shapes as illustrated in Figs.~\ref{Fig3}B and C. The authors in Ref.~\cite{Samueletal1999Berg} examined flagellar filaments with different polymorphic forms, since the different arrangements of the flagellin subunits give rise to grooves with different pitch and chirality~\cite{NambaVonderviszt1997}, as shown in Fig.~{\color{blue}\ref{Fig3}}B.  
	The L-type straight flagellar filament ({\it f0}) that was used by Ref.~\cite{Samueletal1999Berg}  has both left-handed long-pitch grooves and right-handed short-pitch grooves~\cite{BrienBennett1972}. 
	Given that the short-pitch grooves are relevant to the wrapping of the fibres, the findings of Ref.~\cite{Samueletal1999Berg} that bacteria with their flagellar filament in the {\it f0} polymorphic state (i.e.~with right-handed short-pitch grooves) and with a large CCW bias are sensitive to $\chi$-phage  infection, are in agreement with the nut-and-bolt mechanism.
	
	The same study also argued that the translocation time of the phage to the cell body is less than the flagellar filament reversal interval, a necessary condition for successful infection by the virus    of wild-type bacteria whose motors alternate between CCW and CW rotation. Their estimated translocation speeds on the order of microns per seconds give a translocation time which is less than the CCW time interval of about a 1~\rm{s}~\cite{Samueletal1999Berg}.

	Relevant to the nut-and-bolt mechanism are also the findings of an alternative mechanism for adsorption of the flagellotropic phages $\phi$CbK and $\phi$Cb13 that interact with the flagellar filament  of {\it Caulobacter crescentus} using a filament located on the head of the phage~\cite{Guerrero-Ferreira2011}, instead of the tail or tail fibres that other flagellotropic phages use, such as $\chi$ and $PBS1$. This study also reports on  a higher likelihood of infection with a CCW rotational bias that is consistent with the nut-and-bolt mechanism.   
 Notably, phages can also attach to curli fibres, which are  bacterial filaments employed in biofilms; however due to the lack of helical grooves and rotational motion of these filaments, phages are unable to move along them~\cite{Vidakovic2018}. 
	
\begin{figure}[t]
		\includegraphics[width=0.5\columnwidth]{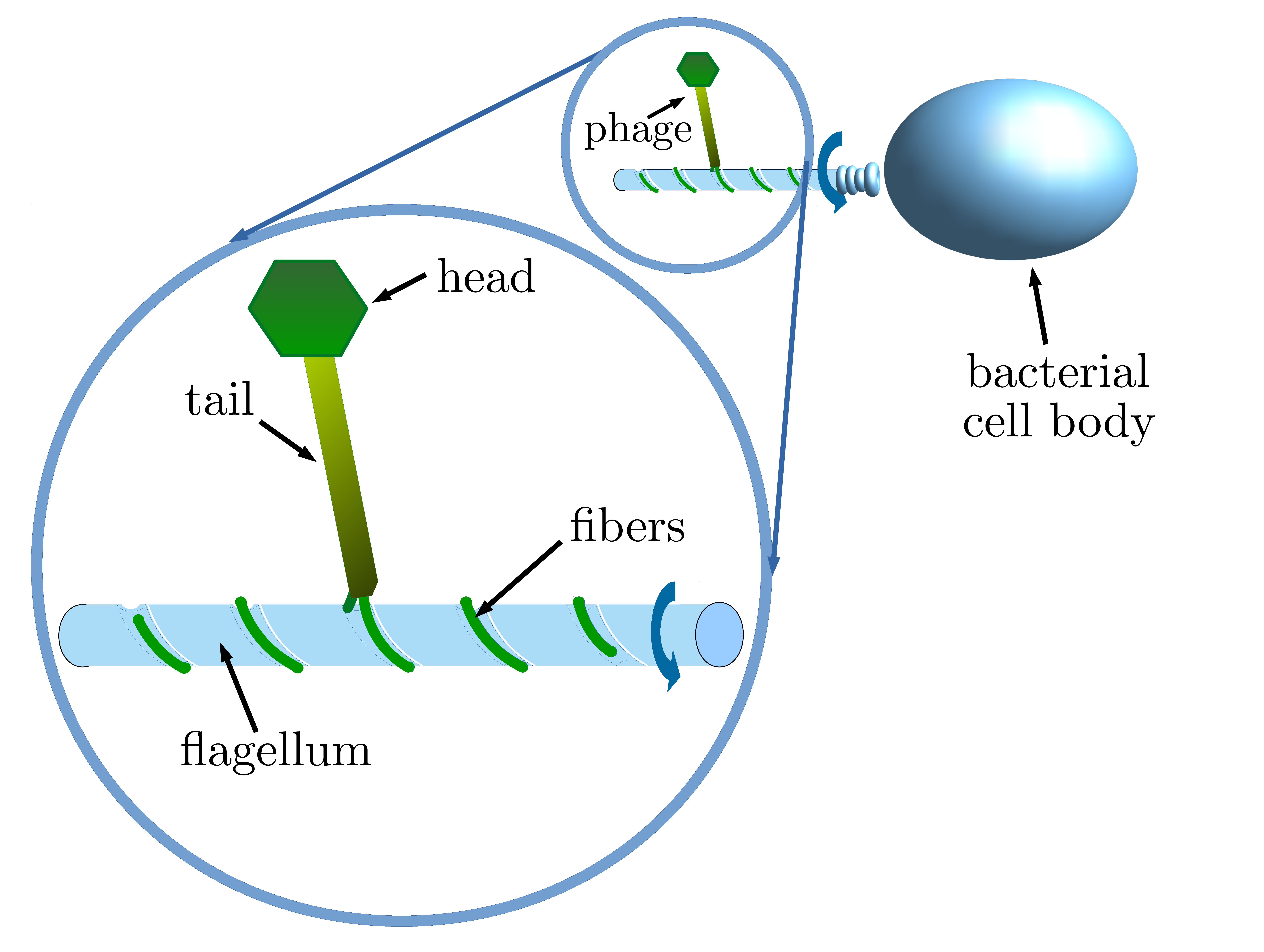}
		\caption{{\bf Schematic  model of the translocation of a flagellotropic phage  along the straight flagellar filament of a mutant bacterium.} The phage, illustrated in dark green, is wrapped around the straight flagellar filament (light blue cylinder) using its fibres, with its tail and head protruding  in the bulk fluid.}\label{Fig4}
	\end{figure}

	In this paper, we theoretically examine the nut-and-bolt mechanism from a quantitative point of view and perform a detailed mathematical analysis of the physical mechanics at play. 
	We focus on  the virus translocation along straight flagellar filaments  in mutants such as the mutant of {\it Salmonella} used in 
	Ref.~\cite{BergAnderson1973}.	
	A flagellotropic phage can wrap around a given flagellar filament using its tail fibres (fibres for short), its tail, or in some cases a filament emanating from the top of its head~\cite{Guerrero-Ferreira2011} and the models we develop can address all these relevant morphologies. 
	A schematic diagram of the typical geometry we consider is shown in Fig.~\ref{Fig4}.
	
	A phage floating in a fluid whose fibres suddenly collide with a flagellar filament rotating at high frequency will undergo a short, transient period of wrapping, during which the length of the fibres that are wrapped around the filament is increasing.
	In this paper we study the translocation of the phage once it has reached a steady, post-wrapping state, and assume that it is moving rigidly with no longer any change in the relative virus-filament configuration.
	
	In order to provide  first-principle theoretical modelling of the nut-and-bolt mechanism, we build in the paper a hierarchy of models. In \S\ref{smooth}, we start with a model of  drag-induced translocation along smooth flagellar filaments that ignores the microscopic mechanics of the grooves yet implicitly captures their effect by coupling the helical shape of the fibres with anisotropy in motion in the local tangent plane of the flagellar filament. Having acquired insight into the key characteristics of the mechanism, we proceed by building a refined, more detailed  model of the guided translocation of phages along grooved flagellar filaments by incorporating the microscopic mechanics of the grooves in \S\ref{grooves}. This is done by including a restoring force that acts to keep the fibres in the centre of the grooves, thereby guiding their motion, as well as a resistive force acting against the sliding motion. In both models, we proceed by considering the geometry, and the forces and torques acting on the different parts of the phage. We use the resistive-force theory of viscous hydrodynamics  in order to model the tail and tail fibres which are both slender~\cite{GrayHancock1955,LaugaPowers2009}.  	 The portion  of the phage wrapping around the flagellar filament is typically the fibres. They experience a hydrodynamic drag from the  motion in the proximity of the rotating flagellar filament along which they slide in the  smooth flagellum model, or a combination of a guiding and resistive forces in the grooved flagellum model. Parts sticking out in the bulk away from the flagellar filament experience a hydrodynamic drag due to their motion in an otherwise stagnant fluid.
	
	We build in our paper a general mathematical formulation relevant to a broad phage morphology. 
	In our typical geometry of phages wrapping around flagellar filaments using their fibres, 
	two limits arise for long-tailed and short-tailed phages.
	Long-tailed phages have their tail and head sticking out in the bulk, away from the flagellar filament, whereas for short-tailed phages only the head is exposed to the bulk fluid. The hydrodynamic torque   actuating the translocation is provided by the parts sticking out in the bulk.
	
	We compare the results from the two models addressing the two geometrical limits and find these to be consistent with each other and with the predictions and experimental observations of Refs.~\cite{BergAnderson1973, Samueletal1999Berg}.  In particular, we predict quantitatively the speed of phage translocation along the flagellar filament they are attached to, and its critical dependence on the interplay between the chirality of the wrapping and the direction of rotation of the filament, as well as the geometrical parameters.
	Most importantly we show that our models capture the correct directionality of translocation, i.e.~that CCW rotation will only pull the phage toward the cell body if the phage slides along a right-handed groove, and predict speeds of translocation on the order of $\mu \rm{m s^{-1}}$, which are crucial for successful infection in the case of bacteria with alternating CCW and CW rotations.  
	
	\section{Drag-induced translocation along smooth flagellar filaments}\label{smooth}
	
	\subsection{Geometry}\label{smooth_geom}	
	
\begin{figure}[t]%[!h]
		\includegraphics[width=0.5\columnwidth]{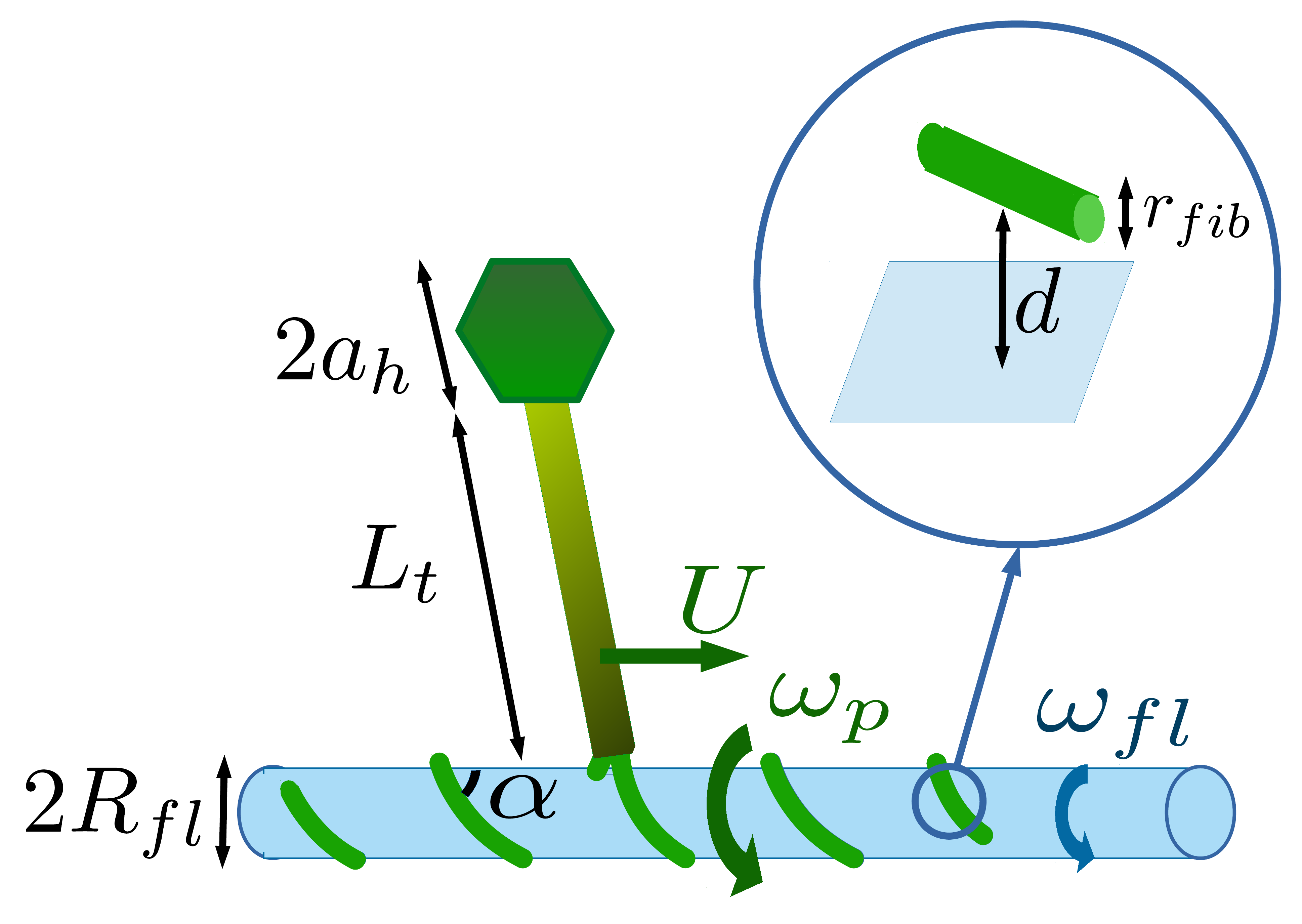}
		\caption{{\bf Mathematical model of drag-induced translocation along a smooth flagellar filament.} The fibres experience an anisotropic  drag due to their motion in the proximity of the rotating flagellar filament. 
			The phage, shown in green, has a capsid head of size $2\ahead$, a tail of length $\Ltail$, fibres of helical shape with helix angle $\angleflag$ and cross-sectional radius $r_{fib}$, and is translocating along a straight flagellar filament, shown in light blue, of radius $\Rflag$. The flagellar filament is rotating at a rate $\omegaflag$. The phage is translocating at speed $U$ and rotating  with rate $\omegaphage$   about the  flagellar filament. The inset shows a short segment of the phage fibre at a distance $d$ from the  local tangent plane of the flagellar filament.}\label{Fig5}
	\end{figure}

	As our first model, we consider the flagellar filament as a straight, smooth rod aligned with the $z$-axis and of radius $\Rflag$. 
	The phage has a capsid head of size $2\ahead$, a tail of length $\Ltail$ and fibres that wrap around the flagellar filament. 	We implicitly capture the effect of the grooves (i) by imposing that the fibres that emanate from the bottom of the tail of the phage are wrapped around the flagellar filament  in a helical shape  and (ii) via the anisotropy in the drag arising from the relative motion between the fibres and the rotating flagellar filament. 
	The helical shape of the fibres has helix angle $\angleflag$, as shown in Fig.~\ref{Fig5}. 
	With the assumption that the gap between the fibres and the flagellar filament is negligible compared to the radius $\Rflag$ of the flagellar filament, the centreline $	\rfib(s)$ of the fibres, parametrised by the contour length position $s$, is described mathematically as
	\begin{align}
	\rfib(s)=\left(\Rflag\cos\left(\frac{s}{\Rflag/\sin\angleflag}\right), h \Rflag\sin\left(\frac{s}{\Rflag/\sin\angleflag}\right), s\cos\angleflag   \right), \quad -\Lfib^{L}<s<\Lfib^{R},
	\end{align}
	with total contour length $\Lfib=\Lfib^{L}+\Lfib^{R}$, where we allow for fibres extending to both sides of the base of the tail to have lengths $\Lfib^{L}$ (left side) and $\Lfib^{R}$ (right side).
	The helix wrapping is  right-handed or left-handed according to whether the chirality index $h$ takes the value $+1$ or $-1$ respectively.
	
	Assuming the phage to move rigidly and working in the laboratory frame, every point $\bv{r}$ on the phage moves with velocity $U\vez + \omegaphage\vez\wedge\bv{r}$.  
	The flagellar filament is assumed to rotate at   rate $\omegaflag$ along its axis, and thus its velocity is given by  $\omegaflag\vez\wedge\bv{r}$ in a fluid that is otherwise stationary, where the value of $\omegaflag$ is known. The purpose of our calculation is to compute the  two unknown quantities, $U$ and $\omegaphage$,  in terms of $\omegaflag$ by enforcing the overall force and torque balance on the phage along the $z$-axis.
	
	\subsection{Forces and moments} \label{smooth_forcesmoments}
	In order to calculate   the forces and torques acting on the tail and fibres we use the 
	resistive-force theory of viscous hydrodynamics (RFT in short)~\cite{GrayHancock1955,LaugaPowers2009}. This theoretical framework  predicts the viscous tractions due to the motion of a slender filament in a viscous fluid by integrating fundamental solutions of the Stokes equations of hydrodynamics~\cite{leal} 
	along the centreline of the filament. 
	
	In an infinite fluid,   the instantaneous hydrodynamic force per unit length exerted on a filament due to its motion in an otherwise stagnant viscous fluid is given by 
	\begin{align}	\bv{f}(s) &=-\resperp\left[\vel(s)-\left(\frac{\partial \bv{r}}{\partial s}\bigg|_s.\vel(s)\right)\frac{\partial \bv{r}}{\partial s}\bigg|_s\right]-\respar \left(\frac{\partial \bv{r}}{\partial s}\bigg|_s.\vel(s)\right)\frac{\partial \bv{r}}{\partial s}\bigg|_s,  \label{RFT}
	\end{align}
	where $\frac{\partial \bv{r}}{\partial s}\big|_s$ and $\bv{v}(s)$ are the local unit tangent and velocity of the filament relative to the fluid at contour-length position $s$ respectively, and  $\respar,\resperp$ are the drag coefficients for motion parallel and perpendicular to the local tangent~\cite{Hancock1953, GrayHancock1955}. For a slender rod of length $L$ and radius $r$ in an infinite fluid, we have 
	\begin{align}
	\resperp\approx \frac{4\pi\mu}{\ln\left({L}/{r}\right)},  \quad \rho\equiv\respar/\resperp\approx1/2,
	\end{align}
	where $\mu$ is the dynamic viscosity of the fluid. 
	
	The fact that the perpendicular drag coefficient is twice the parallel one captures the fact  that it is twice as hard to pull a rod through a viscous fluid in a direction perpendicular to its length than lengthwise. This drag anisotropy is at the heart of the  propulsion physics for  microorganisms such as bacteria and spermatozoa~\cite{LaugaPowers2009}.

	As a result,  the total hydrodynamic force and the $z$-component of the torque on the phage tail due to its motion in the fluid are given by
	\begin{align} 
	\bv{F}_{tail}&=~~~- \int\limits_{0}^{\Ltail}\left[\trespar \ttail \ttail + \tresperp (1 - \ttail \ttail) \right]\cdot \bv{u}^{rel}_{tail}(s) \dd s, \label{FtailRFT}\\
	\vez\cdot\bv{M}_{tail}&=-\vez\cdot \int\limits_{0}^{\Ltail} \rtail(s)\wedge\left\{\left[\trespar \ttail \ttail + \tresperp(1 - \ttail \ttail) \right]\cdot \bv{u}^{rel}_{tail}(s)\right\} \dd s, \label{MtailRFT}
	\end{align}
	where $\rtail(s)$ and $\ttail(s)$ are the position and tangent vectors of the fibre  centreline at contour length position $s$ respectively. The symbols  $\tresperp, \trespar$ are the drag coefficients for motion perpendicular and parallel to the local tangent, with $\trespar\equiv\rhot\tresperp$
	and 
	the velocity of the tail relative to the fluid is
	\begin{align}
	\bv{u}^{rel}_{tail}(s)
	&=\omegaphage\left(\vez\wedge\rtail\right)+ U \vez.
	\label{urelRFTtail}%\\
	\end{align}
	For the fibres, we use the version of RFT modified to capture the motion of slender rods near a surface. The flagellar filament is rotating at rate $\omegaflag$, thus the velocity of the fibres relative to the flagellar filament is given by
	\begin{align}
	\bv{u}^{rel}_{fib}(s)
	&=\Omegarel\left(\vez\wedge\rfib\right)+ U \vez,
	\label{urelRFTfib}
	\end{align}
	where the relative angular velocity is given by
	\begin{equation}
	\Omegarel=\omegaphage   - \omegaflag. 
	\end{equation}

	The expressions for the fibres are similar, and we have
	\begin{align} 
	\bv{F}_{fib}&=~~~
	-\int\limits_{-\Lfib^{(L})}^{\Lfib^{(R)}}\left[\frespar \tfib \tfib + \fresperp(1 - \tfib \tfib) \right]\cdot \bv{u}^{rel}_{fib}(s) \dd s, \label{FfibRFT}\\
	\vez\cdot\bv{M}_{fib}&=-\vez\cdot \int\limits_{-\Lfib^{(L)}}^{\Lfib^{(R)}} \rfib(s)\wedge\left\{\left[\frespar \tfib \tfib + \fresperp(1 - \tfib \tfib) \right]\cdot \bv{u}^{rel}_{fib}(s)\right\} \dd s. \label{MfiblRFT}
	\end{align}
	
	The difference between the expressions in Eqs.~\ref{FtailRFT},\ref{MtailRFT} and Eqs.~\ref{FfibRFT},\ref{MfiblRFT} is that in the latter we use 
	the appropriate resistance coefficients, $\fresperp$ and $\frespar$,   for motion at a small, constant distance $d$ from a nearby surface, in directions perpendicular and parallel to the local tangent of the fibre respectively, given by 
	\begin{align}
	\fresperp\approx \frac{4\pi\mu}{\ln\left({2d}/{r_{fib}}\right)}, 
	\end{align}
	with $\frespar= \rhof \fresperp$ and again $\rhof\approx 1/2$ (see Ref.~\cite{BrennenWinet77} and references therein). 	These results are valid in the limit in which the distance $d$ between the fibre and the surface of the flagellar filaments  is much smaller than the radius of the flagellar filament ($d\ll \Rflag$),  such that the surface of the smooth flagellar filament is locally planar. Importantly, the   very drag anisotropy that allows the rotation of helical flagellar filaments to propel bacteria in the bulk will also enable the rotation of helical fibres around a smooth filament to lead to translocation along the axis of the filament.
	
	If  the tail is straight with total length $\Ltail$ and the head is spherical with  radius $\ahead$, the position of the centre of the head is given by
	\begin{align}
	\rphead=\rbase + \Ltail \ttail + \ahead \thead,
	\end{align}
	where $\rbase=(\Rflag,0,0)$ is the base of the tail from which the fibres emanate. 
	The drag force and torque due to the motion of the head in the otherwise stagnant fluid are given by
	\begin{align}
	\bv{F}_{head}&=- 6\pi \mu \ahead \bv{u}^{rel}_{head}, \label{FheadRFT} \\
	\vez\cdot\bv{M}_{head}
	&=\vez\cdot\left[- 6\pi \mu \ahead  \left(\rphead \wedge\bv{u}^{rel}_{head}\right)\right] - 8\pi\mu\ahead^3 \omegaphage, \label{MheadRFT} 
	\end{align}
	with $\bv{u}^{rel}_{h}$ given by 
	\begin{align}
	\bv{u}^{rel}_{h}
	&=\omegaphage\left(\vez\wedge\rphead\right)+ U \vez.
	\label{urelRFThead}
	\end{align}
	
	Taking $\thead=\ttail$, %=\bv{t}$, 
	the centre of the head will be located  at position
	$\rphead= \rbase + \left(\Ltail + \ahead\right) \ttail$.
	Evaluating the integrals in Eqs.~\ref{FtailRFT},\ref{MtailRFT},\ref{FfibRFT} and \ref{MfiblRFT} and the expressions of Eqs.~\ref{FheadRFT},\ref{MheadRFT} with  this geometry we obtain the forces and torques exerted on the different parts of the phage (projected along the $z$-axis),
	\begin{align}
	\vez\cdot\bv{F}_{fib}&=-\fresperp \Lfib\left[(\sin^2\angleflag+\rho_{fib}\cos^2\angleflag)U -(1-\rho_{fib})\sin\angleflag\cos\angleflag h \Omegarel \Rflag \right],
	\\
	\vez\cdot\bv{M}_{fib}&=- h\fresperp\Rflag\Lfib\left[h\Rflag \Omega_{rel} (\cos^2\angleflag+\rho_{fib}\sin^2\angleflag)-(1-\rho_{fib})\sin\angleflag \cos\angleflag U\right],
	\\
	\vez\cdot\bv{F}_{tail}&=-\tresperp\Ltail\bigg\{ U\left[1-(1-\rhot)\tz^2 \right] - \omegaphage \Rflag(1-\rhot)\ty \tz  \bigg\}, \label{Ftailsmooth}
	\\
	\vez\cdot\bv{M}_{tail}&=-\tresperp\bigg\{ \omegaphage\left[\Ltail\Rflag^2 + \Ltail^2 \Rflag\tx +\frac{1}{3}\Ltail^3 (\tx^2 + \ty^2) \right]\nonumber\\&
	\qquad\qquad\qquad\qquad-(1-\rhot)\ty(U\tz +\omegaphage\Rflag\ty)\Rflag\Ltail \bigg\},
	\\
	\vez\cdot\bv{F}_{head}&=-6\pi\mu\ahead U,
	\\
	\vez\cdot\bv{M}_{head}&=-6\pi\mu\ahead\omegaphage\left[\Rflag
	^2 + 2\Rflag(\Ltail+\ahead)\tx +(\Ltail+\ahead)^2(\tx^2+\ty^2) \right] - 8\pi\mu\ahead^3\omegaphage,
	\end{align}
	where we use the notation $\ttail=(\tx,\ty,\tz)$ for the components of the tangent of the tail.

	\subsection{Phage translocation: General formulation} \label{smooth_genform}
	
	The overall force and torque balance on the phage along the $z$-axis is written as
	\begin{align}
	0&=\vez\cdot[\bv{F}_{fib} + \bv{F}_{tail} + \bv{F}_{head}],  \qquad
	0=\vez\cdot[\bv{M}_{fib} + \bv{M}_{tail} + \bv{M}_{head}],  
	\end{align}
	which leads to  
	\begin{align}
	0=&-\fresperp \Lfib\left[(\sin^2\angleflag+\rho_{fib}\cos^2\angleflag)U -(1-\rho_{fib})\sin\angleflag\cos\angleflag h \Omegarel \Rflag \right]
	\nonumber\\
	&-\tresperp\Ltail\bigg\{ U\left[1-(1-\rhot)\tz^2 \right] - \omegaphage \Rflag(1-\rhot)\ty \tz  \bigg\}
	-6\pi\mu\ahead U,\\
	\	0=&- h\fresperp\Rflag\Lfib\left[h\Rflag \Omega_{rel} (\cos^2\angleflag+\rho_{fib}\sin^2\angleflag)-(1-\rho_{fib})\sin\angleflag \cos\angleflag U\right]
	\nonumber\\
	&-\tresperp\bigg\{ \omegaphage\left[\Ltail\Rflag^2 + \Ltail^2 \Rflag\tx +\frac{1}{3}\Ltail^3 (\tx^2 + \ty^2) \right]
	-(1-\rhot)\ty(U\tz +\omegaphage\Rflag\ty)\Rflag\Ltail \bigg\}
	\nonumber\\
	&-6\pi\mu\ahead\omegaphage\left[\Rflag
	^2 + 2\Rflag(\Ltail+\ahead)\tx +(\Ltail+\ahead)^2(\tx^2+\ty^2) \right] - 8\pi\mu\ahead^3\omegaphage.
	\end{align}
	Writing this system in a formal matrix form for $U$ and $\omegaphage$ in terms of $\omegaflag$ gives
	\begin{align}\label{eq:matrix}
	\begin{pmatrix}A&B\\B&D\end{pmatrix}\begin{pmatrix}U\\\omegaphage\end{pmatrix}=\begin{pmatrix}Z\\H\end{pmatrix}\omegaflag,
	\end{align}
	where we have 
	\begin{align}
	A&=~\tresperp\Ltail\left[1-(1-\rhot)\tz^2\right]+\fresperp\Lfib(\sin^2\angleflag+\rhof\cos^2\angleflag) + 6\pi\mu\ahead, \label{A}\\
	B&=-\left[\tresperp\Ltail(1-\rhot)\ty\tz+h\fresperp\Lfib(1-\rhof)\sin\angleflag\cos\angleflag\right] \Rflag, \label{B}\\
	D&=~\tresperp\Ltail\left[ \Rflag^2 + \Ltail\Rflag\tx + \frac{\Ltail^2}{3}(1-\tz^2)-(1-\rhot)\Rflag^2\ty^2\right] \nonumber\\
	&~~~+\fresperp\Lfib\Rflag^2 (\cos^2\angleflag+\rhof\sin^2\angleflag)\nonumber\\
	&~~~+6\pi\mu\ahead\left[\Rflag^2 + 2\Rflag(\Ltail +\ahead)\tx + (\Ltail +\ahead)^2(1-\tz^2)\right] \nonumber\\
	&~~~+ 8\pi\mu\ahead^3, \label{D}\\
	Z&=-h\fresperp\Lfib \Rflag(1-\rhof)\sin\angleflag\cos\angleflag,\label{Z} \\
	H&=\fresperp\Lfib \Rflag^2 (\cos^2\angleflag + \rhof\sin^2\angleflag)\label{H}.
	\end{align}
	Inverting Eq.~\ref{eq:matrix} gives the translation linear and rotational speeds as
	\begin{align}
	\begin{pmatrix}U\\\omegaphage\end{pmatrix}=\frac{1}{AD-B^2}\begin{pmatrix}D&-B\\-B&A\end{pmatrix}\begin{pmatrix}Z\\H\end{pmatrix}\omegaflag \label{Uexpression}.
	\end{align}
	\normalsize\normalfont
	The full expressions for $(DZ-BH)$,$(AD-BC)$ and $(-CZ+AH)$ for a general phage geometry are given in the Supplementary Material (see~\cite{Supplementary_Material}). %S1 Appendix. % \ref{AppSmooth}\ref{AppSmooth_GenForm}.

	\subsection{Two limits: long vs short-tailed phages} \label{twolimitslongvsshort}
	From the experimental images in  Fig.~\ref{Fig2} we can distinguish   two geometries of wrapping according to how far the tail and head are sticking out in the bulk fluid and away from the flagellar filament.
	We thus proceed by considering the two limiting geometries of long-  and short-tailed phages.
	
	\subsubsection{Long-tailed phages}
	\label{SmoothflagLongLimSection}
	Examples of long-tailed morphology include the $\chi$-phage of {\it E.~coli} and the {\it PBS1} phage of {\it B.~subtilis} shown in Figs.~\ref{Fig2}B and D.
	We use below the $\chi$-phage as a typical long-tailed phage, whose detailed dimensions are reported in Ref.~\cite{SchadeAdlerRis1967}.
	The hexagonal head measures $650-675$~\AA ~between parallel sides (that is $2\ahead\approx650-675$~\AA).
	The tail is a flexible rod that is $2,200$~\AA~long and $140$~\AA~wide,
	and the tail fibres are $2,000-2,200$~\AA~long and $20-25$~\AA~wide. The flagellar filaments are $5-10~\mu \rm{m}$ long and have a diameter $20~\rm{nm}$, hence $\Rflag\approx 100~$\AA~\cite{Chattopadhyayetal2006}. 
	For this specific phage, we thus have $\ahead\approx 30~\rm{nm}$, $\Ltail\approx 220~\rm{nm}$,  $\Lfib\approx 200~\rm{nm}$, $r_{fib} \approx 1 ~\rm{nm}$ and $\Rflag\approx 10 ~\rm{nm}$.
	From this we see that we can safely assume that  $\Rflag,\ahead \ll \Ltail,\Lfib$. Notice however that $\Lfib\approx\Ltail$ and that $\Rflag$ and $\ahead$ are of the same order of magnitude.
	Our variables are thus divided into the short lengthscales of $\ahead$, $\Rflag$ and the long lengthscales of $\Lfib,\Ltail$. With these approximations we obtain the translocation speed as
	\begin{align}
	U_{long}\approx&-h\omegaflag\Rflag(1-\rhof)\sin\angleflag\cos\angleflag ~\curlyG_{long},\label{smooth_Ulong}\\
	\curlyG_{long}=&
	\frac{\fresperp\Lfib 	\begin{bmatrix}	\frac{1}{3}\tresperp\Ltail(1-\tz^2)	+ \left[\tresperp\Rflag\tx	+	6\pi\mu\ahead(1-\tz^2)\right]	\end{bmatrix}}{\frac{1}{3}\tresperp\Ltail(1-\tz^2)\bigg[	\tresperp\Ltail\left[1-(1-\rhot)\tz^2\right]	+\fresperp\Lfib(\sin^2\angleflag+\rhof\cos^2\angleflag)	\bigg]}
	\\
	\approx&
	\frac{\fresperp\Lfib 
	}{\bigg[
		\tresperp\Ltail\left[1-(1-\rhot)\tz^2\right]
		+\fresperp\Lfib(\sin^2\angleflag+\rhof\cos^2\angleflag)
		\bigg]}, \label{Ulongsmooth}
	\end{align}
	\normalsize
	with a relative error of $O\left({\ahead}/{\Ltail},{\ahead}/{\Lfib}, {\Rflag}/{\Ltail}, {\Rflag}/{\Lfib}\right)$. Details of the approximation are given in the Supplementary Material (see~\cite{Supplementary_Material}). %Appendix \ref{AppSmooth}\ref{AppSmooth_Long}.
	
	%%%
	\subsubsection{Short-tailed phages} \label{SmoothflagShortLimSection}
	Phages with very short tails  that use their fibres to wrap around flagellar filaments are equivalent geometrically to phages that use their entire tail for wrapping since in both cases there is a filamentous part of the phage wrapped around the flagellar filament and the head is sticking out in the bulk close to the surface of the filament. 
	For example, the phage OWB that infects $V.~parahaemolyticus$ studied in Ref.~\cite{Zhangetal2016FlagelRotResPhage} and shown in Fig.~\ref{Fig2}A, uses its tail for wrapping.	
	
	In order to avoid any confusion, we will carry out the calculations of this section using the geometry of short-tailed phages, and assume that (i) the tail is negligible and (ii)   the fibres are wrapping around the flagellar filament.   In this case we obtain a translocation speed of  
	\begin{align}
	U_{short}&=	-h \Rflag\omegaflag(1-\rhof)\sin\angleflag\cos\angleflag~\curlyG_{short}, \label{smooth_Ushort}\\
	\curlyG_{short}&=
	\frac{\fresperp\Lfib	
		6\pi\mu\ahead\left[(\Rflag+\ahead\tx)^2 +\ahead^2\ty^2+\frac{4}{3}\ahead^2\right] }
	{		\begin{bmatrix}\rhof\fresperp^2\Lfib^2 \Rflag^2
		+6\pi\mu\ahead\fresperp\Lfib\Rflag^2 (\cos^2\angleflag+\rhof\sin^2\angleflag)
		\\+6\pi\mu\ahead\left[\fresperp\Lfib(\sin^2\angleflag+\rhof\cos^2\angleflag)
		+6\pi\mu\ahead\right]\left[(\Rflag+\ahead\tx)^2 +\ahead^2\ty^2+\frac{4}{3}\ahead^2\right]
		\end{bmatrix}}, \label{Ushortsmooth}
	\end{align}
	\normalsize
	with	details of the calculation given in the Supplementary Material (see~\cite{Supplementary_Material}). %\ref{AppSmooth}\ref{AppSmooth_Short}.
	
	 Note that the results for phages which use their tail to wrap around the flagellar filament can be readily obtained by replacing $\Lfib$ with $\Ltail$ and all relevant quantities in the above result.

	\subsubsection{Interpretation  and discussion of the asymptotic results} \label{smooth_interplimits}
	
	We now interpret and compare the results we obtained in  Eqs.~\ref{smooth_Ulong} and \ref{smooth_Ushort}.	As we now see, our formulae give the correct directionality and speed of translocation in agreement with the qualitative predictions and the experimental data of Ref.~\cite{Samueletal1999Berg}, as well as the requirements for translocation, thereby providing insights to the translocation mechanism.

	Firstly, and most importantly, both results for the translocation speeds in Eqs.~\ref{smooth_Ulong} and \ref{smooth_Ushort}  have the  common factor $-h\Rflag\omegaflag (1-\rhof) \sin\angleflag\cos\angleflag$ which is multiplying the positive dimensionless expressions $\curlyG_{long}$ and $\curlyG_{short}$ respectively. The factor $-h\omegaflag$  gives a directionality for $U$ in agreement with the qualitative prediction  of Ref.~\cite{Samueletal1999Berg} that CCW rotation  will only pull the phage toward the cell body if the phage slides along a right-handed groove. Indeed, our model captures this feature: for right-handed helical wrapping ($h=+1$) and CCW rotation of the flagellar filament when viewing the flagellar filament towards the cell body ($\omegaflag<0$),  the phage moves towards the cell body (i.e.~$U>0$). 
	
	Secondly, the factor $(1-\rhof)$ reveals that translocation requires  
	anisotropy in the friction between the fibres and the surface of the flagellar filament (i.e.~$\rhof\neq 1$).
	We interpret the requirement for anisotropy as an indication of the important role of the grooves in guiding the motion of the fibres. 
	The assumption of a helical wrapping of the fibres coupled with this anisotropy simulates the guiding effect of the grooves in this first model by resisting motion perpendicular to the local tangent of the grooves and promoting motion parallel to it.
	
	Thirdly, the presence of the factor $\sin\angleflag\cos\angleflag$ shows the requirement of a proper helix i.e.~there is no translocation in the  limiting cases of a straight   	 ($\angleflag=0$) or circular wrapping	 	 ($\angleflag=\pi/2$).
	
	Fourthly,   translocation   requires a non-vanishing value of $\Lfib$ in the numerator of both Eqs.~\ref{smooth_Ulong} and \ref{smooth_Ushort}. This is because the fibres are providing the `grip' by wrapping around the flagellar filament.

	Fifthly, the  terms involving the tail and head appear in both the numerator and denominator of Eq.~\ref{smooth_Ulong}. Similarly, terms involving the head appear in both the numerator and denominator of Eq.~\ref{smooth_Ushort}. These show that the parts of the phage that are sticking out in the bulk 	are contributing to both the torque   actuating the motion of the phage relative to the flagellar filament and   the drag. 
	In the case of short phages,  only the head is providing the torque, hence the terms in the numerator of Eq.~\ref{smooth_Ushort}.
	
	Finally, focusing on the $\chi$-phage, the lengths of the tail and the fibres are similar and the logarithmic dependence of the resistance coefficients allow us to estimate the fraction in Eq.~\ref{smooth_Ulong} to be of $O(1)$.  The grooves have a pitch of approximately $50~\rm{nm}$~\cite{Samueletal1999Berg} and the radius of the flagellar filament  is approximately $10~\rm{nm}$, giving rise to helix angle $\angleflag \approx 51^{\circ}$. With $\omegaflag\approx100~\rm{Hz}$, we have that $U=O(\mu \rm{m s^{-1}})$. Importantly, this means that it is possible for  $\chi$-phage to translocate along a flagellar filament of a few $\mu \rm{m}$ long within the timescale of a second, in agreement  with the CCW time interval  for bacteria with alternating CCW and CW rotation, thereby enabling the phage to reach the cell body and infect the bacterium.

	\subsection{Dependence  of translocation speed on geometrical parameters} \label{smooth_geom_dep}
	We now illustrate the dependence of the translocation speed on the geometrical parameters of the phage, namely the lengths $\Ltail$ and $\Lfib$.
	The asymptotic formulae we obtained above and discussed in \S\ref{twolimitslongvsshort} will help verify the asymptotic behaviour of $U$ for large values of $\Ltail$ and   for vanishing tail length, as well as explain the trends for the translocation speed with increasing $\Ltail$ and $\Lfib$.
	
\begin{figure}[t]%[!h]
		\includegraphics[width=0.7\columnwidth]{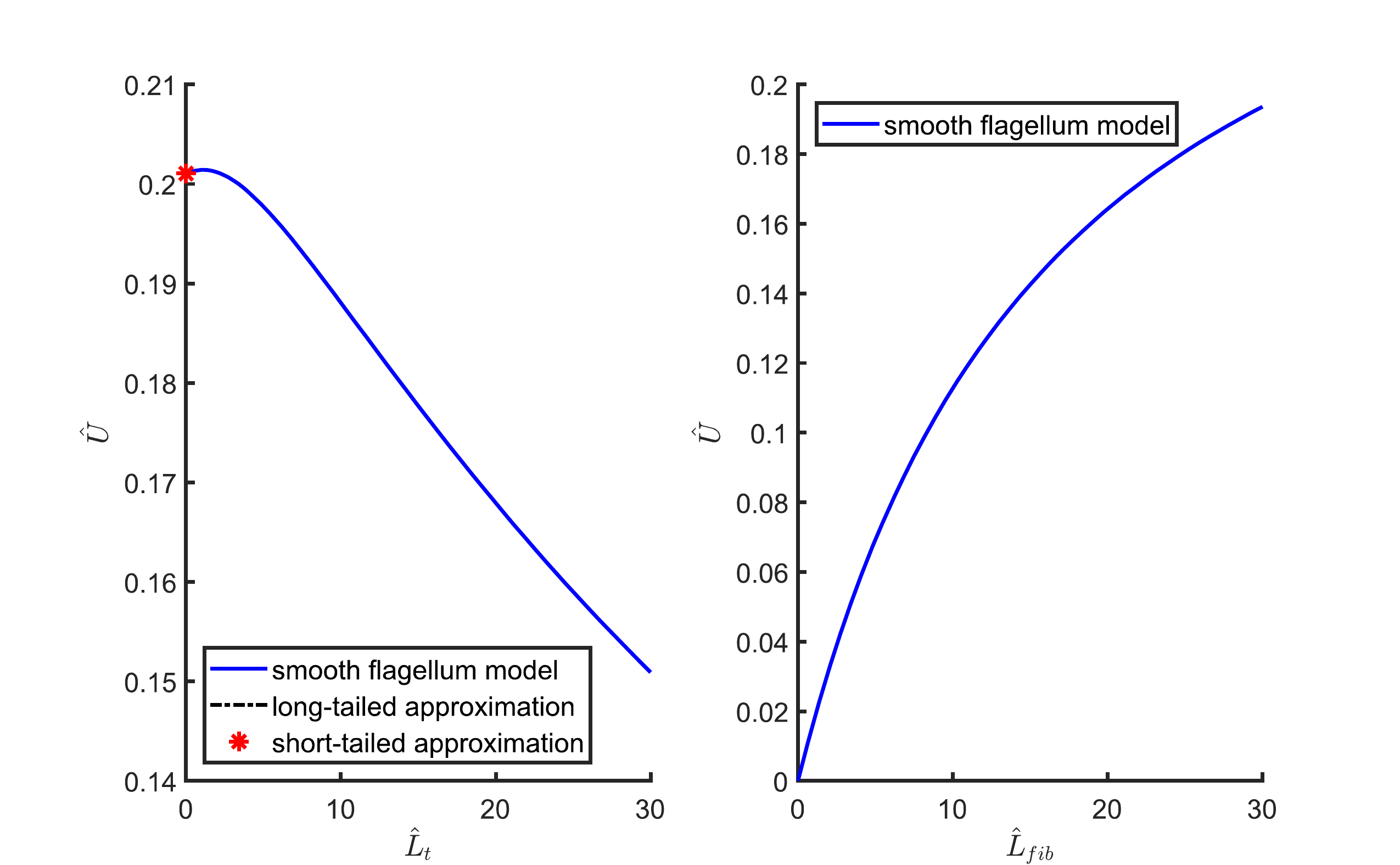}
		\caption{	
			{\bf Smooth flagellum model: Dependence  of the translocation speed, $U$, on the geometrical parameters of the phage.} Left: $U$ vs.~the length of the phage tail, $\Ltail$, based on Eq.~\ref{Uexpression} (blue solid line). The speed is a decreasing function of $\Ltail$.   The value of $U$ for vanishing tail length is well captured by the theoretical approximation for short-tailed phages (red star). For large values of $\Ltail$ it approaches the theoretical approximation for long-tailed phages (black dash-dotted line, inset). 
			Right: $U$ vs.~the length of the fibres, $\Lfib$,  following  Eq.~\ref{Uexpression}. The speed is an increasing function of $\Lfib$.  The long-tailed approximation of Eq.~\ref{Ulongsmooth} also captures this increasing behaviour.}		\label{U_Phage_Translocation_Smooth}
	\end{figure}

	To fix ideas, we consider the specific case of the $\chi$ phage and hence the following set of parameter values (as in \S\ref{SmoothflagLongLimSection} and Ref.~\cite{SchadeAdlerRis1967}):  $\ahead\approx 30~\rm{nm}$, $\Ltail\approx 220~\rm{nm}$,   $r_{tail}=7~\rm{nm}$, $\Lfib\approx 200~\rm{nm}$, $r_{fib} \approx 1 ~\rm{nm}$,  $\Rflag\approx 10 ~\rm{nm}$ and $\mu=10^{-3}~\rm{Pa.s}$. For the helix angle we take $\angleflag=51^{\circ}$ (see \S\ref{smooth_interplimits}). We also take $(\tx,\ty,\tz)=(1,1,1)/\sqrt{3}$.
	Note that for simplicity we do not include the slow variation of the resistive coefficients with $\Ltail$, but instead  keep their constant non dimensional values, $\tresperphat=4\pi/\ln(220/7)$, and $\fresperphat=4\pi/\ln(4)$
	based on $\tresperp\approx 4\pi\mu/\ln\left({\Ltail}/{r_{tail}}\right)$ with $\Ltail/r_{tail}=220/7$ and $\fresperp\approx 4\pi\mu/\ln\left({2d}/{r_{fib}}\right)$ with $d=2 r_{fib}$.
	We then use Eq.~\ref{Uexpression} to plot $U$ versus $\Ltail$ and $\Lfib$ in Fig.~\ref{U_Phage_Translocation_Smooth}. For simplicity we have non-dimensionalised lengths by $\Rflag$, time by $\omegaflag^{-1}$ and viscosity by $\mu$, and denote dimensionless quantities using a hat.

	In Fig.~\ref{U_Phage_Translocation_Smooth}A we observe that the  phage translocation   speed is a decreasing function of $\Ltail$.   The value of $U$ for vanishing tail length is well captured by the theoretical approximation for short-tailed phages  of Eq.~\ref{Ushortsmooth} (red star).  	For large values of $\Ltail$, it approaches the theoretical approximation for long-tailed phages (black dash-dotted line, inset). For the long-tailed approximation we used the expressions in Eq.~A1-A2 in the Supplementary Material (see~\cite{Supplementary_Material}) keeping terms up to and including $L^3$, for $L$ either $\Ltail$ or $\Lfib$. We note that the smaller the phage head size, $\ahead$, the better the convergence between   Eq.~\ref{Uexpression} and the long-tailed approximation (which assumes $\ahead, \Rflag\ll \Ltail,\Lfib$). In Fig.~\ref{U_Phage_Translocation_Smooth}A we used the dimensions for the $\chi$ phage, that correspond to $\ahead/\Lfib=3/22$, and the long-tailed approximation also requires $ \aheadhat \gg \Ltailhat$.
	The long-tailed approximation from Eq.~\ref{Ulongsmooth} captures the decreasing behaviour of $U$ with $\Ltail$ and can be used to explain this result. Specifically, the term involving $\Ltail$ in the denominator of Eq.~\ref{Ulongsmooth}  shows that this decay arises from the resistive part of the hydrodynamic force on the tail (the first term in Eq.~\ref{Ftailsmooth}) which increases as $\Ltail$ increases. In other words, when $\Ltail$ increases the drag increases and therefore the speed decreases.

	In Fig.~\ref{U_Phage_Translocation_Smooth}B we next show the speed $U$ as a function of the length of the fibres, $\Lfib$. Clearly the speed is an increasing function of $\Lfib$ and  the long-tailed approximation of Eq.~\ref{Ulongsmooth} captures this increasing behaviour. We observe that	terms involving $\Lfib$ appear in both the numerator and denominator of Eq.~\ref{Ulongsmooth}. 
	Dividing top and bottom by $\Lfib$, we see that as $\Lfib$ increases, $U$ increases (at some point $U$ will asymptote to a constant value, but the length at which this happens  appears to be too large to be relevant biologically). The increase of $U$ with $\Lfib$ stems from the  propulsive forces per unit length on the fibres that integrate to a larger propulsive torque as $\Lfib$ increases. Of course there is also the resistive drag on the fibres but that starts to become more important only at larger values of $\Lfib$.

	\section{Guided translocation along grooved flagellar filaments} \label{grooves}
	
	\subsection{Geometry}\label{grooves_geom}
	As a more refined physical model, we now include in this section the   mechanics arising from the microscopic details of the grooved surface of the flagellar filament due to the packing of the flagellin molecules and	modify the previous calculation in order  to account for the motion of the phage fibres sliding along the helical grooves.
	
\begin{figure}[t]%[!h]
		\includegraphics[width=0.5\columnwidth]{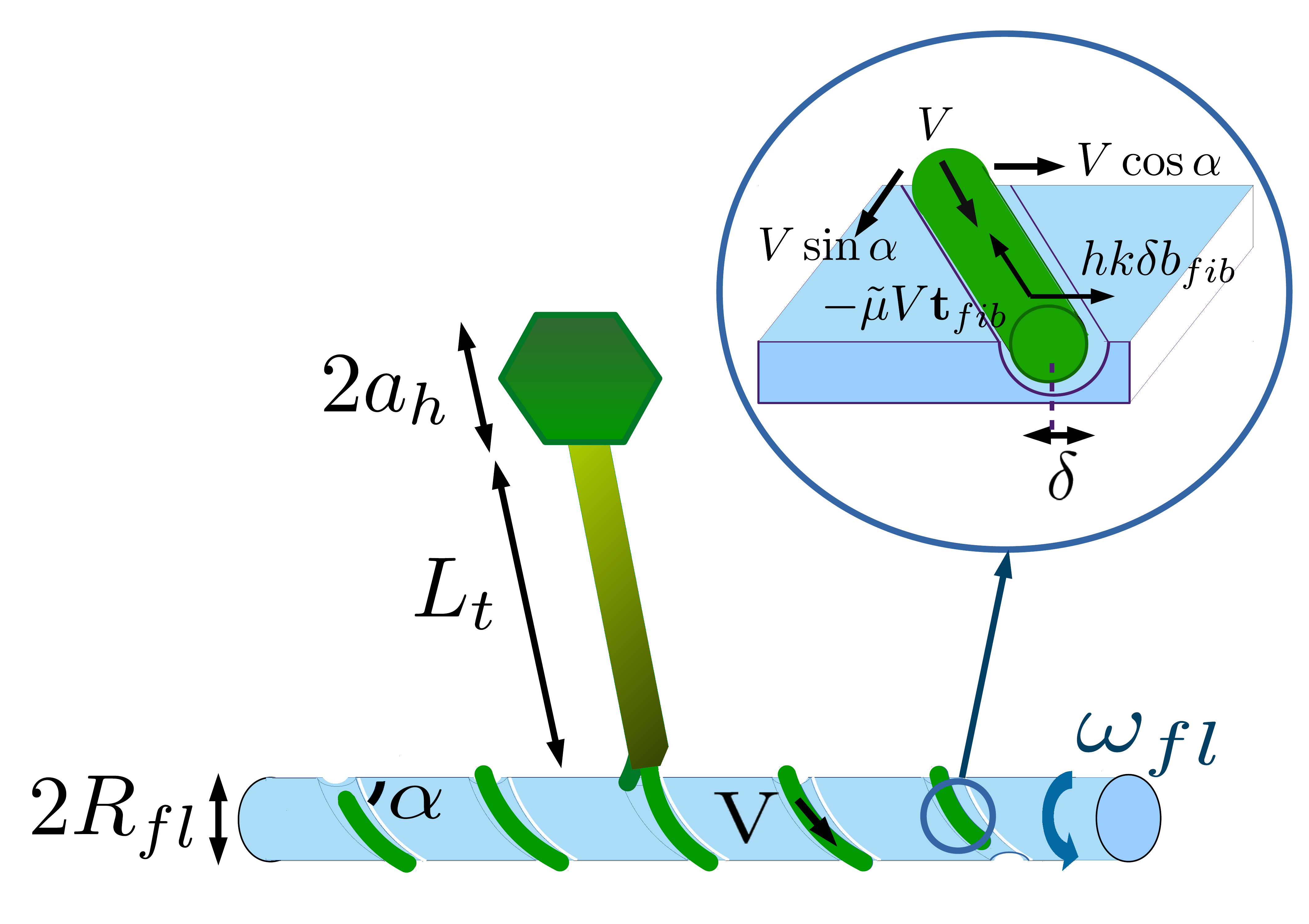}
		\caption{{\bf Guided translocation of phage along a grooved flagellar filament.} 
			The phage, shown in green, has a capsid head of size $2\ahead$, a tail of length $\Ltail$ and fibres of cross-sectional radius $r_{fib}$.
			The flagellar filament (light blue) has helical grooves of helix angle $\angleflag$ and is rotating at a rate $\omegaflag$. The phage slides along the grooves with speed $V$ in the frame of the flagellar filament.
			As shown in the inset, the force acting on the fibre sliding along the grooves consists of two parts: (i) a drag resisting the sliding motion of magnitude $\mutil V$ in the $-\tfib$ direction and (ii) a restoring force acting to keep the fibre in the centre of the groove of magnitude $\kdpot $ in the $h\bfib$ direction, where $\dpot$ is the local offset of the centre of the fibre cross-section from the centre of the groove and $\bfib$ is the local binormal to the fibre centreline that lies in the local tangent plane of the surface of the flagellar filament and is perpendicular to the tangent of the fibre  centreline.
		}\label{Figgroovesmodel}
	\end{figure}

	If the phage slides with speed V along the grooves of helix angle $\angleflag$ in the frame of the straight flagellar filament, as illustrated in Fig.~\ref{Figgroovesmodel}, then the translocation velocity and rotation rate measured in the laboratory frame,  $U$ and $\omegaphage$, become
	\begin{align}
	U&=V\cos\angleflag,\\
	\omegaphage&=\frac{hV\sin\angleflag}{\Rflag}+ \omegaflag.
	\end{align}

	With this substitution we obtain the forces and torques acting on the tail and head as
	\begin{align}
	\vez\cdot\bv{F}_{tail}&=-\tresperp\Ltail\bigg\{ U\left[1-(1-\rhot)\tz^2 \right] - \omegaphage \Rflag(1-\rhot)\ty \tz  \bigg\},
	\\
	\vez\cdot\bv{M}_{tail}&=-\tresperp\bigg\{ \omegaphage\left[\Ltail\Rflag^2 + \Ltail^2 \Rflag\tx +\frac{1}{3}\Ltail^3 (\tx^2 + \ty^2) \right]\nonumber\\
	&\qquad\qquad\qquad\qquad
	-(1-\rho_{t})\ty(U\tz +\omegaphage\Rflag\ty)\Rflag\Ltail \bigg\}, \label{Mtailgrooves}
	\\
	\vez\cdot\bv{F}_{head}&=-6\pi\mu\ahead U,
	\\
	\vez\cdot\bv{M}_{head}&=-6\pi\mu\ahead\omegaphage\left[\Rflag
	^2 + 2\Rflag(\Ltail+\ahead)\tx +(\Ltail+\ahead)^2(\tx^2+\ty^2) \right] - 8\pi\mu\ahead^3\omegaphage.
	\end{align}
	
	\subsection{Forces and moments} \label{grooves_forcesmoments}
	The details of the interactions between the phage fibres and the grooves are expected to be complicated as they   depend on the parts of the flagellin molecules that make up the groove surface and interact with the proteins that the fibres consist of.  These interactions could originate from a number of short range    intermolecular forces, for example electrostatic repulsion or Van der Waals forces. 
	We model here the resultant of the interaction forces acting on the fibre sliding along the grooves as consisting of two parts, a drag and a restoring force, as shown in the inset of Fig.~\ref{Figgroovesmodel}.
	
	Firstly, the fibre is subject to a viscous drag,  $-\mutil V\tfib$ per unit length where $\mutil$ is a hydrodynamic resistance coefficient against  the sliding motion (with dimensions of a viscosity).
	A simple approximation for that coefficient is to assume that there is a fully-developed  shear flow resisting the sliding between the fibres and the surface of the grooves. Assuming the cross-section of the latter to be a circular arc, so that a fraction $f_{cov}$ of the circumference of a cross-section of the fibres lies inside the groove, we obtain approximately  
	\begin{equation}
	\mutil\approx\frac{2\pi r_{fib} f_{cov} \mu}{\hgap},
	\end{equation}
	where $\hgap$ is the size of the gap  between the grooves and the fibres and $r_{fib}$ is the radius of the fibres. 	
	
	Secondly, there should be a restoring force 
	acting to keep the fibre in the centre of the groove arising from the physical interactions between the fibre and the groove.  	A simple modelling approach consists of viewing each  side of the groove as repelling the  fibre, with the resultant of these forces providing a restoring force 
	$h\kdpot\bfib(s)$ per unit length, arising from a potential well $\frac{1}{2}k\dpot^2$ where $\dpot$ is the distance from the centre of the well, and $\bfib$ is the local binormal vector to the fibre centreline,
	\begin{align}
	\bfib=\bigg[ h\cos\alpha\sin\left(\frac{s}{\Rflag/\sin\angleflag}\right),-\cos\alpha\cos\left(\frac{s}{\Rflag/\sin\angleflag}\right), h\sin\alpha\bigg],
	\end{align}
	that lies in the local tangent plane of the surface of the flagellar filament and is perpendicular to the tangent vector $\tfib$ of the fibre  centreline.
	Assuming $\dpot$ to be uniform along the length of the fibres, the expressions for the force and torque on the fibres given as the integrals,
	\begin{align}
	\bv{F}_{fib}&= \int_{-\Lfib^{(L)}}^{\Lfib^{(R)}}-\mutil V\tfib(s) - h\kdpot\bfib(s) \dd s ,
	\\
	\bv{M}_{fib}&=\int_{-\Lfib^{(L)}}^{\Lfib^{(R)}}\rfib(s)\wedge\left[-\mutil V\tfib(s) - h\kdpot\bfib(s)\right] \dd s ,
	\end{align}
	when projected along the $z$ direction become
	\begin{align}
	\vez\cdot\bv{F}_{fib}&=\left[-\mutil V\cos\angleflag -\kdpot\sin\angleflag\right]\Lfib,\\
	\vez\cdot\bv{M}_{fib}&=-\left[\mutil V h\Rflag\sin\angleflag - h \kdpot \Rflag\cos\angleflag\right]\Lfib.
	\end{align}

	\subsection{Phage translocation: General formulation} \label{grooves_genform}
	
	The total force and torque balance on the phage along the $z$-axis,
	\begin{align}
	0&=\vez\cdot[\bv{F}_{fib} + \bv{F}_{tail} + \bv{F}_{head}], 
	\qquad 0=\vez\cdot[\bv{M}_{fib} + \bv{M}_{tail} + \bv{M}_{head}],  
	\end{align}
	give the system to be solved in order to find the two unknown quantities, $V$ and $k\delta$, in terms of $\omegaflag$,
	\begin{align}
	0=&\left[-\mutil V\cos\angleflag -\kdpot\sin\angleflag\right]\Lfib
	\nonumber\\
	&-\tresperp\Ltail\bigg\{ V\cos\angleflag\left[1-(1-\rhot)\tz^2 \right] - \left(\frac{hV\sin\angleflag}{\Rflag}+ \omegaflag\right) \Rflag(1-\rhot)\ty \tz  \bigg\}
	\nonumber\\
	&-6\pi\mu\ahead V\cos\angleflag, \label{grooveforcebal}\\
	0=&-\left[\mutil V h\Rflag\sin\angleflag -h\kdpot \Rflag\cos\angleflag\right]\Lfib
	\nonumber\\
	&-\tresperp\bigg\{ \left(\frac{hV\sin\angleflag}{\Rflag}+ \omegaflag\right)\left[\Ltail\Rflag^2 + \Ltail^2 \Rflag\tx +\frac{1}{3}\Ltail^3 (\tx^2 + \ty^2) \right]\nonumber\\
	&\qquad\qquad\qquad\qquad
	-(1-\rho_{t})\ty\left[V\cos\angleflag\tz +\left(\frac{hV\sin\angleflag}{\Rflag}+ \omegaflag\right)\Rflag\ty\right]\Rflag\Ltail \bigg\}
	\nonumber\\
	&-6\pi\mu\ahead\left(\frac{hV\sin\angleflag}{\Rflag}+ \omegaflag\right)\left[\Rflag
	^2 + 2\Rflag(\Ltail+\ahead)\tx +(\Ltail+\ahead)^2(\tx^2+\ty^2) \right] \nonumber\\&~~- 8\pi\mu\ahead^3\left(\frac{hV\sin\angleflag}{\Rflag}+ \omegaflag\right).\label{groovetorquebal}
	\end{align}
	
	Using matrix notation, these two equations take the form
	%
	%%%%%%
	\begin{align}\label{eq:matrix2}
	\begin{pmatrix}A&B\\C&D\end{pmatrix}\begin{pmatrix}V \\\kdpot\end{pmatrix}=\begin{pmatrix}Z\\H\end{pmatrix}\omegaflag,
	\end{align}
	where
	\begin{align}
	A&=~\bigg\{ \tresperp\Ltail\left(   
	\cos\angleflag\left[1-(1-\rhot)\tz^2 \right]
	-h\sin\angleflag(1-\rhot)\ty \tz
	\right)+\mutil \Lfib\cos\angleflag +6\pi\mu\ahead \cos\angleflag\bigg\}, \label{Agrooves}\\
	B&=~\Lfib\sin\angleflag, \label{Bgrooves}\\
	C&=~\bigg\{h\mutil  \Lfib \Rflag\sin\angleflag 
	+\tresperp \frac{h\sin\angleflag}{\Rflag}\left[\Ltail\Rflag^2 + \Ltail^2 \Rflag\tx +\frac{1}{3}\Ltail^3 (\tx^2 + \ty^2) \right]\nonumber\\
	&\qquad
	-(1-\rho_{t})\ty\tresperp \left[\cos\angleflag\tz +h\sin\angleflag \ty\right]\Rflag\Ltail 
	\nonumber\\
	&\qquad
	+6\pi\mu\ahead\frac{h\sin\angleflag}{\Rflag}\left[\Rflag
	^2 + 2\Rflag(\Ltail+\ahead)\tx +(\Ltail+\ahead)^2(\tx^2+\ty^2) \right]+8\pi\mu\ahead^3\frac{h\sin\angleflag}{\Rflag}
	\bigg\}, \label{Cgrooves}\\
	D&=~-h \Rflag \Lfib\cos\angleflag, \label{Dgrooves}\\
	Z&=~\tresperp\Ltail \Rflag(1-\rhot)\ty \tz, \label{Zgrooves}\\
	H&=~-\bigg\{\tresperp\left[\Ltail\Rflag^2 + \Ltail^2 \Rflag\tx +\frac{1}{3}\Ltail^3 (\tx^2 + \ty^2)-(1-\rho_{t})\ty^2 \Rflag^2\Ltail \right]
	\\	
	&\qquad\qquad+6\pi\mu\ahead \left[\Rflag
	^2 + 2\Rflag(\Ltail+\ahead)\tx +(\Ltail+\ahead)^2(\tx^2+\ty^2) \right] +8\pi\mu\ahead^3\bigg\}. \label{Hgrooves}
	\end{align}
	The details of the calculation are given in the Supplementary Material (see~\cite{Supplementary_Material}). % \ref{AppGrooves}\ref{AppGrooves_GenForm}.
	
	Inverting Eq.~\ref{eq:matrix2}, $V$ and $\kdpot$ are obtained as 	\begin{align}
	\label{eq:V}	V&=\frac{DZ-BH}{AD-BC}\omegaflag,
	\\ \kdpot&=\frac{-CZ+AH}{AD-BC}\omegaflag,
	\end{align}
	where the full expressions for $(DZ-BH)$,$(AD-BC)$ and $(-CZ+AH)$ for a general phage geometry are given in the Supplementary Material (see~\cite{Supplementary_Material}). % \ref{AppGrooves}\ref{AppGrooves_GenForm}. 	
	Finally, from Eq.~\ref{eq:V}, the translocation velocity along the $z$-axis is calculated as $U=V\cos\angleflag$.

	\subsection{Two limits: long vs short-tailed phages} \label{twolimitslongvsshortGroove}
	We now proceed by considering the two limiting geometries of long-tailed and short-tailed phages similarly to  
	\S\ref{twolimitslongvsshort}.
	\subsubsection{Long-tailed phages} \label{GroovesLongLimSection} 
	Under the approximations relevant for long-tailed phages such as $\chi$-phage described in \S\ref{twolimitslongvsshort}, i.e.~$\Rflag,\ahead \ll \Ltail,\Lfib$, 
	the translocation velocity along the $z$-axis gets simplified to
	\begin{align}
	U_{long}&=-h\Rflag\omegaflag\sin\angleflag\cos\angleflag\curlyG_{long}
	,
	\label{grooves_Ulong} \\
	\curlyG_{long}&=\frac{
		\Ltail^2 \left[\frac{1}{3}\tresperp\Ltail
		+6\pi\mu\ahead\right](1-\tz^2) }
	{\Ltail^2 \left[\frac{1}{3}\tresperp\Ltail
		+6\pi\mu\ahead\right](1-\tz^2)\sin^2\angleflag
		+\mutil \Rflag^2\Lfib }\cdot \label{grooves_Glong}
	\end{align}
	with the details of the approximation   given in the Supplementary Material (see~\cite{Supplementary_Material}). %Appendix~\ref{AppGrooves}\ref{AppGrooves_Long}.

	\subsubsection{Short-tail phages}
	\label{GroovesShortLimSection} 
	In the case of short-tail phages, we assume that the tail is negligible and that the fibres are wrapping around the flagellar filament. The translocation velocity simplifies then to
	\begin{align}
	U_{short}&=-h\Rflag\omegaflag\sin\angleflag\cos\angleflag\curlyG_{short},\label{grooves_Ushort} \\
	\curlyG_{short}&=\frac{
		6\pi\mu\ahead \left[\Rflag
		^2 + 2\Rflag\ahead\tx +\ahead^2(\frac{7}{3}-\tz^2) \right]}{\mutil \Lfib\Rflag^2 
		+6\pi\mu\ahead\left[\Rflag^2
		+ 2\Rflag\ahead\tx\sin^2\angleflag +\ahead^2\sin^2\angleflag\left(\frac{7}{3}-\tz^2 \right)\right]}\cdot \label{grooves_Gshort}
	\end{align}
	with all calculation  details in the Supplementary Material (see~\cite{Supplementary_Material}).

	\subsubsection{Interpretation and discussion of the results}
	Similarly to \S\ref{smooth_interplimits}, we interpret and compare the results in Eqs.~\ref{grooves_Ulong} and \ref{grooves_Ushort}. Here again, the crucial factor $-h\Rflag\omegaflag\sin\angleflag\cos\angleflag$ appears in both equations multiplying a positive, non-dimensional expression, and  we obtain the correct directionality and speed of translocation in agreement with Ref.~\cite{Samueletal1999Berg}. The prefactor $-h\omegaflag$ gives the correct directionality for $U$, i.e.~for right-handed helical wrapping ($h=+1$)
	and CCW rotation of the flagellar filament ($\omegaflag<0$),  the phage moves towards the cell body ($U>0$), in agreement with Ref.~\cite{Samueletal1999Berg}. 
	Again, the translocation speed of $O(\mu\rm{m}/s)$ allows translocation during the CCW time interval for bacteria that alternate between CCW and CW sense, in agreement with Ref.~\cite{Samueletal1999Berg}.  
	The factor $\sin\angleflag\cos\angleflag$ shows   that a  proper helix is needed  for translocation.
	The presence of the term $\mutil\Rflag^2\Lfib$ in the denominator implies that the sliding drag from the fibre decreases the translocation speed, and longer fibres give a decreased speed. Further, and similarly to \S\ref{smooth_interplimits}, the terms inside the square brackets in both the numerator and denominator of Eqs.~\ref{grooves_Glong} and \ref{grooves_Gshort} show that the parts of the phage that are sticking out in the bulk (for the long phages these are the tail and the head, for the short phages it is only the head), are contributing to both the torque that is actuating the motion of the phage relative to the flagellar filament and to the drag.

	\subsection{Dependence  of   translocation speed on   geometrical parameters} \label{groove_geom_dep}

	\begin{figure}[t]%[t]
		\includegraphics[width=0.8\columnwidth]{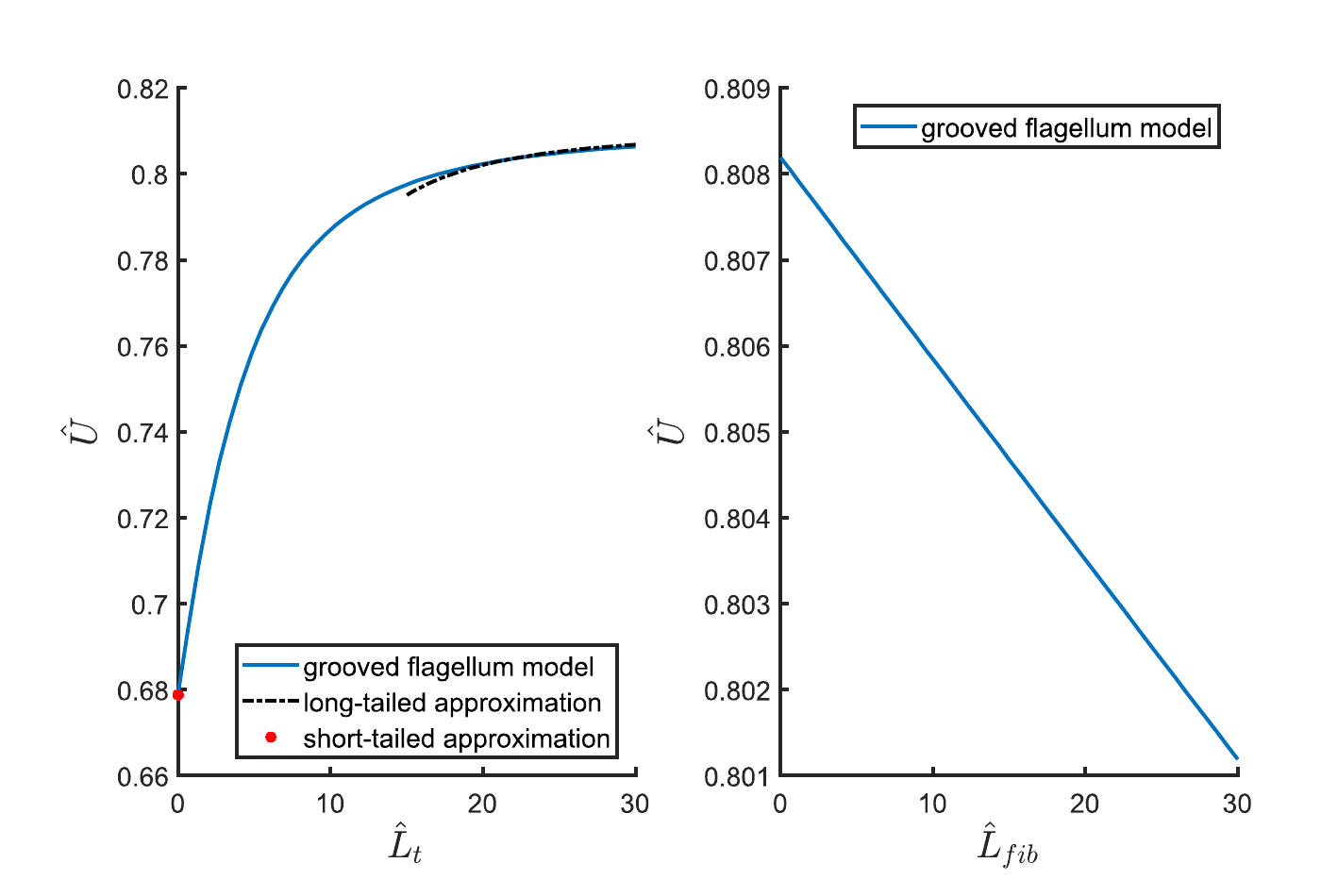}
		\caption{	
			{\bf Grooved flagellum model: Dependence  of the translocation speed on the geometrical parameters of the phage.} Left:   $U$ as a function of the length of the phage tail, $\Ltail$, based on Eq.~\ref{eq:V} (blue solid line). The speed is an increasing function of $\Ltail$.   The value of $U$ at vanishing tail length is well captured by the theoretical approximation for short-tailed phages (red star). For large values of $\Ltail$, it approaches the theoretical approximation for long-tailed phages (black dash-dotted line).
			Right:   $U$ vs.~the length of the fibres, $\Lfib$,  based on Eq.~\ref{eq:V}.
			The long-tailed approximation of Eq.~\ref{grooves_Glong} captures the decreasing behaviour of $U$ with $\Lfib$.			}	
		\label{U_Phage_Translocation_Groove}
	\end{figure}

	We now illustrate the dependence of the translocation speed on the geometrical parameters of the phage, namely $\Ltail$ and $\Lfib$, according to our model of translocation along grooved flagellar filaments.
	We use the same approach, parameter values and non-dimensionalisation as in \S\ref{smooth_geom_dep} and as there   we denote dimensionless quantities using a hat. For $\hat{\mutil}\equiv\mutil/\mu\approx2\pi r_{fib} f_{cov} /\hgap$, we take $r_{fib}/h_{gap}=2$, i.e.~assume the gap between the grooves and the fibres is half the fibre cross-sectional radius, and $f_{cov}=1/2$, so that the cross-section of the grooves is a semi-circle. These parameter values give $\hat{\mutil}=2\pi$.
	We then use Eq.~\ref{eq:V} to plot $\hat{U}$ (calculated as $\hat{U}=\hat{V}\cos\angleflag$) as a function of both  $\Ltailhat$ and $\Lfibhat$ in Fig.~\ref{U_Phage_Translocation_Groove}.
 	In Fig.~\ref{U_Phage_Translocation_Groove}A we observe that the translocation  speed is an increasing function of $\Ltail$.   The value of $U$ at vanishing tail length is well captured by the theoretical approximation for short-tailed phages  of Eq.~\ref{grooves_Gshort} (red star in figure). For large values of $\Ltail$, the result  approaches the theoretical approximation for long-tailed phages (black dash-dotted line).

	The long-tailed approximation from Eq.~\ref{grooves_Glong}, through the terms with  $\Ltail$ in both numerator and denominator,  is able to  capture the increasing behaviour of $U$ with $\Ltail$.  
	Physically, this trend is caused by the propulsive terms in $\vez\cdot\bv{M}_{tail}$ in Eq.~\ref{Mtailgrooves} (proportional to $\Ltail^3$) that increase as $\Ltail$ increases.
 	Next in Fig.~\ref{U_Phage_Translocation_Groove}B we show the speed $U$ as a function of the length of the fibres, $\Lfib$,  as predicted by   Eq.~\ref{eq:V}, and observe a decreasing trend.
	The long-tailed approximation of Eq.~\ref{grooves_Glong} is able to capture this   behaviour. 
	The presence of the term involving $\Lfib$ in the denominator of Eq.~\ref{grooves_Glong} leads to a decrease of $U$ with $\Lfib$, and is physically due to an increase of the viscous drag on the fibres as $\Lfib$ increases.

\begin{figure}[t]%[t]
		\includegraphics[width=0.5\columnwidth]{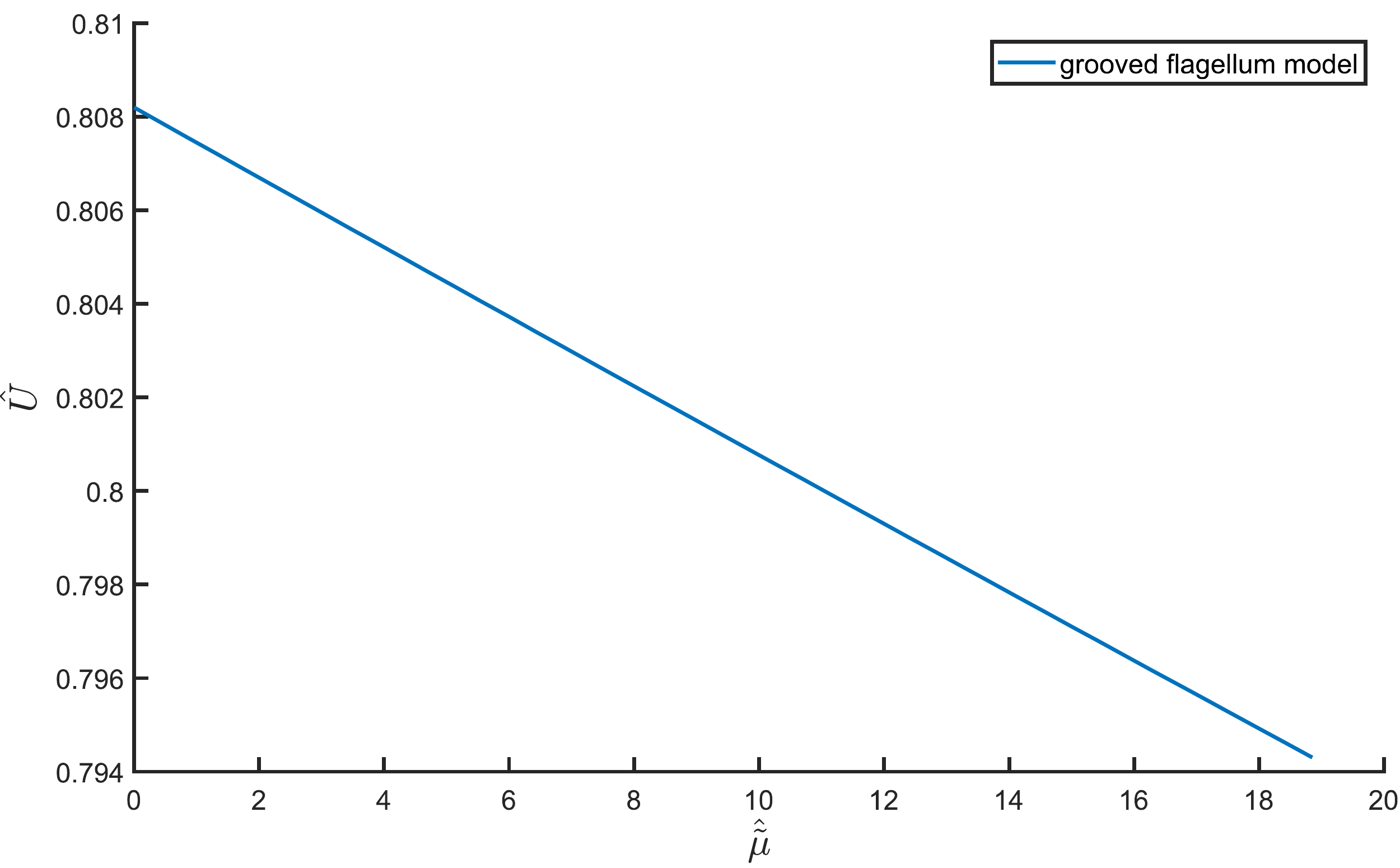}
		\caption{
			{\bf Grooved flagellum model: Dependence  of the phage translocation speed, $U$, on the `effective' viscosity, $\hat{\mutil}$ (non-dimensionalised by the fluid viscosity $\mu$).}  
			As $\mutil$ increases, the  resistive part of the force from the motion of the fibres in the grooves increases, which slows down the phase and leads to a decrease of $U$.}	
		\label{Fig_Grooves_U_mutil_PLots}
	\end{figure}

	Finally, as shown in Fig.~\ref{Fig_Grooves_U_mutil_PLots}, we obtain that the translocation speed is a decreasing function of the  `effective’ viscosity in the grooves, $\hat{\mutil}$, due to  the resistive part of the force from the motion of the fibres in the grooves, as expected.

	\section{Conclusion}\label{conclusion}
	In this work, we carried out a first-principle theoretical study of the nut-and-bolt mechanism of phage translocation along the straight flagellar filaments of bacteria. 
	The main theoretical predictions from our two models,  
	Eqs.~\ref{smooth_Ulong}, \ref{smooth_Ushort}, \ref{grooves_Ulong} and \ref{grooves_Ushort}, give the phage translocation speed, $U$,  in terms of the phage and groove geometries and the rotation rate of the flagellar filament, in the two relevant limits of long- and short-tailed phages. These mathematical results capture the basic qualitative experimental observations and predictions of Refs.~\cite{BergAnderson1973, Samueletal1999Berg} for the speed and directionality of translocation which are both crucial for successful infection. 
	
	The common prefactor in the formulae for the translocation speed along the filament, $U \sim -h \omegaflag \Rflag \sin\angleflag\cos\angleflag$, appears in the expressions from both models. This provides the expected directionality in agreement with Refs.~\cite{BergAnderson1973, Samueletal1999Berg}:  sliding of the fibres along flagellar filaments with right-handed helical grooves ($h=+1$), combined with CCW rotation of flagellar filament $(\omegaflag<0)$, will give rise to phage translocation towards the cell body ($U>0$) for infection to follow, whereas CW rotation ($\omegaflag>0$) would translocate the phage away from the cell body ($U<0$), towards the free end of the flagellar filament, and thus away from the cell body.
	
	Quantitatively, the speeds predicted by our model are estimated to a few micrometres per second,  $U=O(\mu \rm{m s^{-1}})$. This is important for phages infecting bacteria which alternate between CCW and CW senses of rotation. The phage needs translocation speeds of this magnitude in order to move along a flagellar filament of a few $\mu \rm{m}$  long within a timescale of $1~\rm{s}$, which is approximately the CCW time interval~\cite{Samueletal1999Berg}.
	
	Furthermore, the two limits of long- and short-tailed phages clarify that the physical requirements for translocation along the flagellar filament are the grip from the part of the phage that is wrapped around the flagellar filament, in a helical shape (indeed as dictated by the shape of the grooves), combined with torque provided by the parts of the phage sticking out in the bulk, away from the flagellar filament. 
	The plots for the translocation speed as a function of the phage tail length approach the  asymptotic approximations for long- and short- tailed phages at large and vanishing tail lengths respectively in both models.
	
	The important point where the two models deviate from each other is their opposite predictions for the     translocation speed as a function of the phage tail length and the phage fibre length.  We conjecture that the second model with its explicit inclusion of the grooves should be closer to the real-life situation. According to it, $U$ increases when $\Ltail$ increases because the propulsive terms in the axial torque on the tail increase with $\Ltail$. In contrast, the decreasing behaviour of $U$ with $\Lfib$ is caused by the resistive part of the force exerted by the motion of the fibres in the grooves that increases with $\Lfib$, and thus slows down the motion.

	Having modelled phage translocation along straight flagellar filaments of mutant bacteria, the next natural step will be to model phage translocation along the naturally helical flagellar filaments of wild type bacteria. 
	In this case, the geometry is more complicated as it involves motion along helical grooves on top of a flagellar filament whose centreline is also a helix. 
	This is a more complicated system geometrically, as the helical fibres are sliding along a helical flagellar filament with a spatially-varying local tangent  and therefore we expect that  numerical computations would be  required in order to tackle it. 	Notably, there will be an additional hydrodynamic drag on the phage due to the rotation and translation of the flagellar filament. The hydrodynamic drag from the translation will oppose or enhance the translocation of the phage towards the cell body depending on the chirality of the flagellar filament.  
	This opens up the possibility of a competition between 
	the nut-and-bolt translocation effect and the possibly opposing drag due to translation, which will vary with the helical angle of the flagellar filament. Different regimes are expected to arise as the helical angle of the flagellar filament is increased from $0$, hinting to a rich, nonlinear behaviour to be investigated with the aid of numerical simulations.

	In this work, we focus on the 
	translocation of the phage once it has reached a steady, post-wrapping state, and thus assuming that it is moving rigidly.
	Future studies could address   the 
	transient period of wrapping, where the length of the fibres   wrapped around the filament is increasing and the `grip' is possibly becoming tighter. Some phages (in the {\it Siphoviridae} family)  
	have long, flexible tails, thereby requiring the addition of the elasticity of the tail and fibres into the model.  We hope that the   modelling developed in this paper will motivate not only further theoretical studies along those lines  but also more experimental work  clarifying   the  processes involved in the wrapping and motion of the     fibre  in the grooves.

	\section*{Acknowledgements}
	This work was funded by the EPSRC (PK) and by the European Research Council (ERC) under the European Union's Horizon 2020 research and innovation programme (grant agreement 682754 to EL).

	\bibliography{NutnBoltRef}
\end{document}